\documentclass{bmcart}

\usepackage{amsthm,amsmath}
\usepackage{bbm}
\usepackage{MnSymbol}
\RequirePackage{hyperref}
\usepackage[utf8]{inputenc} 
\usepackage{amsfonts}
\usepackage{graphicx}
\usepackage{subcaption}
\usepackage{array,booktabs}
\usepackage{multirow}
\usepackage{algorithm}
\usepackage{algpseudocode}
\usepackage{afterpage}

\newtheorem{theorem}{Theorem}
\newtheorem{lemma}{Lemma}
\newtheorem{proposition}{Proposition}
\newtheorem{corollary}{Corollary}
\newtheorem{definition}{Definition}
\newtheorem{remark}{Remark}

\newcolumntype{M}[1]{>{\centering\arraybackslash}m{#1}}

\DeclareMathOperator*{\argmax}{arg\,max}
\DeclareMathOperator*{\argmin}{arg\,min}

\startlocaldefs
\endlocaldefs

\begin{document}

\begin{frontmatter}

\begin{fmbox}
\dochead{Research}


\title{$L^{\gamma}$-PageRank for Semi-Supervised Learning}


\author[
   addressref={aff1,aff2},                   
   corref={aff1},                       
   email={esteban.bautista-ruiz@ens-lyon.fr}   
]{\inits{EB}\fnm{Esteban} \snm{Bautista}}
\author[
   addressref={aff2},
   email={patrice.abry@ens-lyon.fr}
]{\inits{JRS}\fnm{Patrice} \snm{Abry}}
\author[
   addressref={aff1},
   email={paulo.goncalves@ens-lyon.fr}
]{\inits{JRS}\fnm{Paulo} \snm{Gon\c{c}alves}}


\address[id=aff1]{
  \orgname{Univ Lyon, Inria, CNRS, ENS de Lyon, UCB Lyon 1, LIP UMR 5668}, 
  \postcode{F-69342}                                
  \city{Lyon},                              
  \cny{France}                                    
}
\address[id=aff2]{%
  \orgname{Univ Lyon, Ens de Lyon, Univ Claude Bernard, CNRS, Laboratoire de Physique},
  \postcode{F-69342}
  \city{Lyon},
  \cny{France}
}


\begin{artnotes}
\end{artnotes}

\end{fmbox}


\begin{abstractbox}

\begin{abstract} 
PageRank for Semi-Supervised Learning has shown to leverage data structures and limited tagged examples to yield meaningful classification. Despite successes, classification performance can still be improved, particularly in cases of fuzzy graphs or unbalanced labeled data. To address such limitations, a novel approach based on powers of the Laplacian matrix $L^\gamma$ ($\gamma > 0$), referred to as $L^\gamma$-PageRank, is proposed.
Its theoretical study shows that it operates on signed graphs, where nodes belonging to one same class are more likely to share positive edges while nodes from different classes are more likely to be connected with negative edges. It is shown that by selecting an optimal $\gamma$, classification performance can be significantly enhanced. A procedure for the automated estimation of the optimal $\gamma$, from a unique observation of data, is devised and assessed. Experiments on several datasets demonstrate the effectiveness of both $L^\gamma$-PageRank classification and the optimal $\gamma$ estimation.

\end{abstract}




\end{abstractbox}
%

\end{frontmatter}




\section{Introduction}  

\subsection{Context}
Graph-based Semi-Supervised Learning (G-SSL) is a modern important tool for classification. While Unsupervised Learning fully relies on the data structure and Supervised Learning demands extensive labeled examples, G-SSL combines limited tagged examples and the data structure to provide satisfactory results. This makes the field of G-SSL of utmost importance as nowadays large and structured datasets can be readily accessed in comparison to expert data which may be hard to obtain. Examples where G-SSL provide state of the art results are vast, ranging from classification of BitTorrent contents and users \cite{Avrachenkov.2012.classification}, text categorization \cite{Subramanya.2008.soft}, medical diagnosis \cite{Zhao.2014.compact}, or zombie hunting under BGP protocol \cite{Fontugne.2019.BGP}. Algorithmically, PageRank constitutes the reference tool in G-SSL. It has spurred a deluge of theory \cite{Chung.2010.pagerank, Avrachenkov.2018.mean, Litvak.2009.characterization, Chung.2007.four}, applications \cite{Avrachenkov.2008.pagerank, Graham.2009.distributing, Avrachenkov.2012.classification, Fontugne.2019.BGP} and implementations \cite{Andersen.2007.using, Andersen.2007.detecting}. Despite successes, the performance of G-SSL can still be improved, particularly for fuzzy graphs or imbalance of labeled datasets, two situations that we aim to address in this work.

\subsection{Related works}
In graphs, a ground truth class is represented by a subset of graph nodes, denoted $S_{gt}$. Thus, in graphs, the classification challenge corresponds to finding the binary partition of the graph vertices: $\mathcal{V} = S_{gt}\cup S_{gt}^c$. If the data is structured, then $S_{gt}$ forms a cluster, i.e., a densely and strongly connected graph region that is weakly connected to the rest of the graph. This is exploited by G-SSL methods that essentially amount to diffuse information placed on the tagged nodes of $S_{gt}$, through the graph, expecting a concentration of information in $S_{gt}$ that reveals its members. Among the family of G-SSL propositions obeying this rationale \cite{Zhou.2003.learning, Zhou.2007.spectral, Avrachenkov.2012.generalized}, PageRank is considered the state of the art approach in terms of performance, algorithms and theoretical understanding. The PageRank algorithm can be interpreted as random walkers that start from the labeled points and, at each step, diffuse to an adjacent node with probability $\alpha$ or restart to the starting point with probability $(1 - \alpha)$. In the limit (of infinite steps), each node is endowed a score proportional to the number of visits to it. Thus, vertices of $S_{gt}$ are expected to get larger scores as walkers get trapped for a long time by the connected structure of $S_{gt}$. The capacity of PageRank to confine the random walks within $S_{gt}$ depends on a topological parameter of $S_{gt}$ known as the Cheeger ratio, or conductance, counting the ratio of external and internal connections of $S_{gt}$. More precisely, it is shown in \cite{Andersen.2007.using} that the probability of a PageRank random walker leaving $S_{gt}$ is upper bounded by the Cheeger ratio of $S_{gt}$. In other terms, a small Cheeger ratio designates a strongly disconnected cluster that PageRank can eventually easily detect. Based on the scores, a binary partitioning via a sweep-cut procedure allows to retrieve an estimate $\hat{S}_{gt}$. This procedure is granted to obtain an estimate $\hat{S}_{gt}$ with a small Cheeger ratio if a sharp drop in magnitude appears on the sorted scores, then $\hat{S}_{gt}$ is potentially a good estimation of the ground truth $S_{gt}$ \cite{Andersen.2007.detecting}. In \cite{Zhou.2011.semi}, an issue affecting G-SSL methods, coined as the `curse of flatness', was highlighted. Such work proposes to extend PageRank by iterating the \emph{random walk} Laplacian in the PageRank solution, as a mean to enforce Sobolev regularity to the vertex scores and amend the aforementioned problem. However, with this approach, guarantees that a sweep-cut still leads to a meaningful clustering remains unproven and it can be given neither diffusion nor topological interpretations. Thus, preventing insights on the properties and qualities of partitions it retrieves. This makes it hard to build upon and to address the issues listed above.

\subsection{Goals, contributions and outline}
In this work, we revisit Laplacian powers as a way to improve G-SSL and to address the issues listed above. We propose a generalization  of PageRank by using (non necessarily integers) powers of the \emph{combinatorial} Laplacian matrix $L^\gamma$ ($\gamma > 0$). In contradistinction to \cite{Zhou.2011.semi}, our approach (i) enables us to have an explicit closed form expression of the underlying optimization problem (see Eq. \ref{LgPR-optimization.eq}); (ii) permits a diffusion and a topological interpretation. In our approach, we show that, for each $\gamma$, a new graph is generated. These new graphs, which we refer to as $L^\gamma$-graphs, reweight the links of the original structure and create edges, which can be positive or negative, between initially far-distant nodes. This topological change has the potential to improve classification as the signed edges introduce what can be seen as agreements (positive edges) or disagreements (negative edges) between nodes, allowing to revamp clusters as groups of nodes agreeing between them and disagreeing with the rest of the graph. This paper investigates the potential of these $L^\gamma$-graphs to better delineate a targeted $S_{gt}$, compared to PageRank. The theoretical analysis of our proposition permits to extend the Cheeger ratio to $L^\gamma$-graphs and to prove that if there is a $L^{\gamma}-$graph in which $S_{gt}$ has a smaller Cheeger ratio, then we can more accurately identify it with our generalized $L^{\gamma}$-PageRank procedure using the sweep-cut technique. Then, by means of numerical investigations, we point the existence of an optimal $\gamma$ value that maximizes performance. Finally, we propose an algorithm that allows to estimate the optimal $\gamma$ directly from the graph and the labeled points.

The paper is organized as follows: Section \ref{soa.sec} sets definitions and recalls classical results on G-SSL. Section \ref{LgPR.sec} presents the main contributions of the paper: Section \ref{Lg-topology.sec} introduces $L^\gamma$-graphs; Section \ref{LgPR-method.sec} defines $L^\gamma$-PageRank and its theoretical analysis; Section \ref{g-tunig.sec} discusses the existence of an optimal $\gamma$ and its estimation. Section \ref{Performance.sec} shows the improvements in classification performance permitted by $L^\gamma$-PageRank on several real world datasets commonly used in classification, as well as the relevance of the estimation procedure for the optimal tuning.

\section{State of the art}\label{soa.sec}
\subsection{Preliminaries}
Let $\mathcal{G}(\mathcal{V},\mathcal{E}, w)$ denote a weighted undirected graph with no self-loops in which: $\mathcal{V}$ refers to the set of vertices of cardinality $|\mathcal{V}| = N$; $\mathcal{E}$ denotes the set of edges, where a connected pair $u,v \in \mathcal{V}$, denoted $u \sim v$, implies $(u,v), (v,u) \in \mathcal{E}$; and $w : \mathcal{E} \to \mathbb{R}^{+}$ is a weight function. The graph adjacency matrix is denoted by $W$ in which $W_{uv} = w(u,v)$ if $u \sim v$ and $W_{uv} = 0$ otherwise. For a vertex $u \in \mathcal{V}$ we let $d_u = \sum_v W_{uv}$ denote the degree of $u$ and $D = diag(d_1, \dots, d_N)$ be the diagonal matrix of degrees. Let $\Delta_{uv}$ denote the geodesic distance between $u$ and $v$. Given a set of nodes $S \subseteq \mathcal{V}$, we denote by $\mathbbm{1}_S$ the indicator function of such set, meaning that $\left({\mathbbm{1}_S}\right)_u = 1$ if $u \in S$ and $\left({\mathbbm{1}_{S}}\right)_u = 0$ otherwise. The volume of $S$ is defined to be $vol(S) = \sum_{u \in S} d_u$. We refer to the volume of the entire graph by $vol(G)$. Let $f : \mathcal{V} \to \mathbb{R}$ be a signal lying on the graph vertices. Graph signals are represented as column vectors, where $f_u$ refers to the signal value at node $u$. The sum of signal values in the set $S$ is denoted by $f(S) = \sum_{u \in S} f_u$. We denote by $L = D - W$ the combinatorial graph Laplacian which, by construction, is a real symmetric matrix with eigendecomposition of the form $L = Q \Lambda Q^T$. The positivity of the Dirichlet form $f^T L f = \sum_{u \sim v} W_{uv}(f_i - f_j)^2 \geq 0$ implies that $L$ has real non-negative eigenvalues.

A random walk on a graph is a Markov chain where the nodes form the state space. Thus, when a walker is located at a node $u$ at a specific time $t$, at time step $t + 1$ the walker moves to a neighbor $v$ with probability $P_{uv}$, where $P = D^{-1}W$. If the graph signal $\chi$ represents the distribution for the random walk starting point, then the signal $ x^T = \chi^T P^t$ denotes the distribution of the walker position at time $t$. Independently of the starting distribution, if the graph is connected and not bipartite, the random walk converges to a stationary distribution $\pi^T = \pi^T P$, where $\pi_u = d_u/vol(G)$.

Clustering is the search of groups of nodes that are strongly connected between them and weakly connected to the rest of the graph. The Cheeger ratio is a metric that counts the ratio of external and internal connections of a group of nodes, thus assessing its pertinence as a cluster, while penalizing uninteresting solutions that may fit the cluster criteria, like isolated nodes linked by a few edges. It is defined as follows.
\begin{definition}
	For a set of nodes $S \subseteq \mathcal{V}$, the Cheeger ratio, or conductance, of $S$ is defined as:
	\begin{equation}
		h_S := \frac{ \sum_{u \in S} \sum_{v \in S^c} W_{uv} }{ \min \{ vol(S), vol(S^c) \} }.
	\end{equation}
\end{definition}
Thus, we define clustering as finding the binary partition of the graph vertices: $\mathcal{V} = S \cup S^c$ such that $S$ has low $h_S$.

\subsection{PageRank-based Semi-Supervised Learning}
Let $\mathcal{V}_{S_{gt}} \subseteq S_{gt}$ denote the set of nodes tagged to belong to the ground truth $S_{gt}$ and $y$ be indicator function of $\mathcal{V}_{S_{gt}}$, i.e. $y_u = 1$ if node $u \in \mathcal{V}_{S_{gt}}$ and $y_u = 0$ otherwise. The PageRank G-SSL is defined as the solution to the optimization problem \cite{Avrachenkov.2012.generalized}:
\begin{equation}\label{PR-optimization.eq}
	\argmin_{f} \left\{ f^T D^{-1}L D^{-1} f + \mu \left(f - y\right)^T D^{-1} \left(f-y \right)\right\}.
\end{equation}

Optimization problem (\ref{PR-optimization.eq}) can be seen as the search of a smooth graph signal in the sense that strongly connected nodes should have similar values (left term), while the labeled data is respected (right term), and a regularization parameter $\mu$ tunes the trade off between both terms. Notably, problem (\ref{PR-optimization.eq}) is convex with closed form solution given by \cite{Avrachenkov.2012.generalized}:
\begin{equation}\label{PR-solution.eq}
	f = \mu \left( LD^{-1} + \mu \mathbb{I} \right)^{-1} y.
\end{equation}
We present the PageRank solution in this form as it will simplify derivations in the reminder of the paper, but it is not hard to rewrite (\ref{PR-solution.eq}) to its more popular version: $f^T =(1 - \alpha) \sum_{k = 0}^{\infty} \alpha^k y^T P^k$ where $\alpha = 1 / (1 + \mu)$. This latter helps to expose the connection between PageRank and diffusion processes. Namely, it corresponds to the equilibrium state of a random walk that decides either to continue with probability $\alpha$, or to restart to the starting distribution $y$ with probability $(1 - \alpha)$. As $y = \sum_{u \in \mathcal{V}_{S_{gt}}} \delta_u$ is the combination of different starting distributions, it is clear that the PageRank score at a particular node is proportional to the probability of finding a walker, at equilibrium, at this node. PageRank diffusion satisfies the following properties \cite{Tsiatas.2012.diffusion}: (\textit{i}) mass preservation: $\sum_{u \in \mathcal{V}} f_u  = \sum_{u \in \mathcal{V}} y_u$; (\textit{ii}) stationarity: $f = \pi$ if $y = \pi$; and (\textit{iii}) limit behavior: $f \to \pi$ as $\mu \to 0$ and $f \to y$ as $\mu \to \infty$. 

In \cite{Andersen.2007.using}, it is shown that the behavior of this type of random walks is tightly related to the cluster structure of graphs. This connection between PageRank and clustering is quantified in the following result. 
\begin{lemma}{\cite{Andersen.2007.using}}\label{PR-bound.lemma}
	Let $S \subset \mathcal{V}$ be an arbitrary set with $vol(S) \leq vol(G)/2$. For a labeled point placed at a node $u \in S$ selected with probability proportional to its degree in $S$, i.e. $d_u/vol(S)$, the PageRank satisfies
\begin{equation}
	\mathbb{E} [ f(S^c) ] \leq \frac{h_S}{\mu}.
\end{equation}
\end{lemma}
This lemma implies that if we apply PageRank diffusion to the labels of $S_{gt}$ and it has a small $h_{S_{gt}}$, then the probability of finding a walker outside $S_{gt}$ is small and the nodes with largest PageRank value should index $S_{gt}$. This is formalized in \cite{Andersen.2007.using} and \cite{Andersen.2007.detecting}. The former shows that a proxy $\hat{S}_{gt}$ that has small $h_{\hat{S}_{gt}}$ can be found by looking for regions of high concentration of PageRank mass. The latter improves that result, showing that $\hat{S}_{gt}$ can be found more easily by looking for a sharp drop in the PageRank scores. To state their result, we first introduce the sweep-cut technique.
\begin{definition}
A sweep-cut is a procedure to retrieve a partition $\mathcal{V} = \hat{S}_{gt} \cup \hat{S}^c_{gt}$ from the PageRank vector. The procedure is as follows: 
\begin{itemize}
    \item Let $v_1, \dots, v_N$ be a rearrangement of the vertices in descending order, so that the permutation vector $q$ satisfies $q_{v_i} = f_{v_i}/d_{v_i} \geq q_{v_{i+1}} = f_{v_{i+1}}/d_{v_{i+1}}$
	\item Let $S_j = \left\{ v_1, \dots, v_j \right\}$ be the set of vertices indexed by the first j elements of $q$.
	\item Let $\tau(f) = \min_j h_{S_j}$
	\item Retrieve $\hat{S}_{gt} = S_{j}$ for the set $S_j$ achieving $\tau(f)$
\end{itemize}
\end{definition}
Now, we state the result of \cite{Andersen.2007.detecting}, showing that if there is a sharp drop in rank at $S_j$, then the set $S_j$ has small Cheeger ratio.
\begin{lemma}{\cite{Andersen.2007.detecting}}
Let $h \in (0,1)$, $j$ be any index in $[1, N]$ and $\alpha \in (0,1]$ denote the PageRank restarting probability. Let $C(S_j,S_j^c) = \sum_{u \in S_j} \sum_{v \in S_j^c} W_{uv}$ be the numerator of the Cheeger ratio. Then, $S_j$ satisfies one of the following: (a) $C(S_j,S_j^c) < 2hvol(S_j)$; or (b) there is some index $k > j$ such that $vol(S_k) \geq vol(S_j)(1 + h)$ and $q_k \geq q_j - \alpha/h vol(S_j)$  
\end{lemma}

In other words, this lemma implies that either $S_j$ has a small Cheeger ratio, or there is no sharp drop at $q_j$. 

\subsection{Generalization to multiple classes}\label{soa_multiClass.sec}
PageRank G-SSL can be readily generalized to a multi-class setting in which labeled points of $K$ classes are used to find a partition $\mathcal{V} = S_{1} \cup S_2 \cup \dots \cup S_{K}$. Let $\mathcal{V}_{S_k}$ denote the labeled points of class $k$ and the indicator function of $\mathcal{V}_{S_k}$ be placed as the $k$-th column of a matrix $Y$. Then, the multi-class PageRank is computed in matrix form as \cite{Sokol.2014.graph}: $\min_{F} \{ F^T D^{-1}L D^{-1} F + \mu \left(F - Y\right)^T D^{-1}\left(F - Y \right)\}$, with classification matrix given in closed form by $F = \mu \left( LD^{-1} + \mu \mathbb{I} \right)^{-1} Y$. This leads a node $u$ to have $K$ associated scores and it is assigned to the cluster $k$ satisfying $\argmax_k F_{uk}$. In \cite{Sokol.2014.graph}, the following rule explaining the classification is provided: let $pr_{uv}$ denote the probability that a random walk reaches node $v$ before restarting to node $u$, then $v$ is assigned to the class $k$ that satisfies the inequality
\begin{equation} 
\sum_{u \in \mathcal{V}_k} pr_{uv} \geq \sum_{w \in \mathcal{V}_{k'}} pr_{wv}, ~\hspace{5pt} \forall k' \neq k.
\end{equation}
This inequality highlights an important issue of the multi-class approach as the sums depend on the cardinality of the sets of labeled points. Thus, cases of unbalanced number of labeled points can potentially bias the classification.

\section{$L^{\gamma}$-PageRank for Semi-Supervised Learning}\label{LgPR.sec}


\subsection{The $L^{\gamma}$-graphs}\label{Lg-topology.sec}
\begin{figure}[t]
\begin{subfigure}[t]{0.45\textwidth}
\centering
    \includegraphics[width=0.9\textwidth]{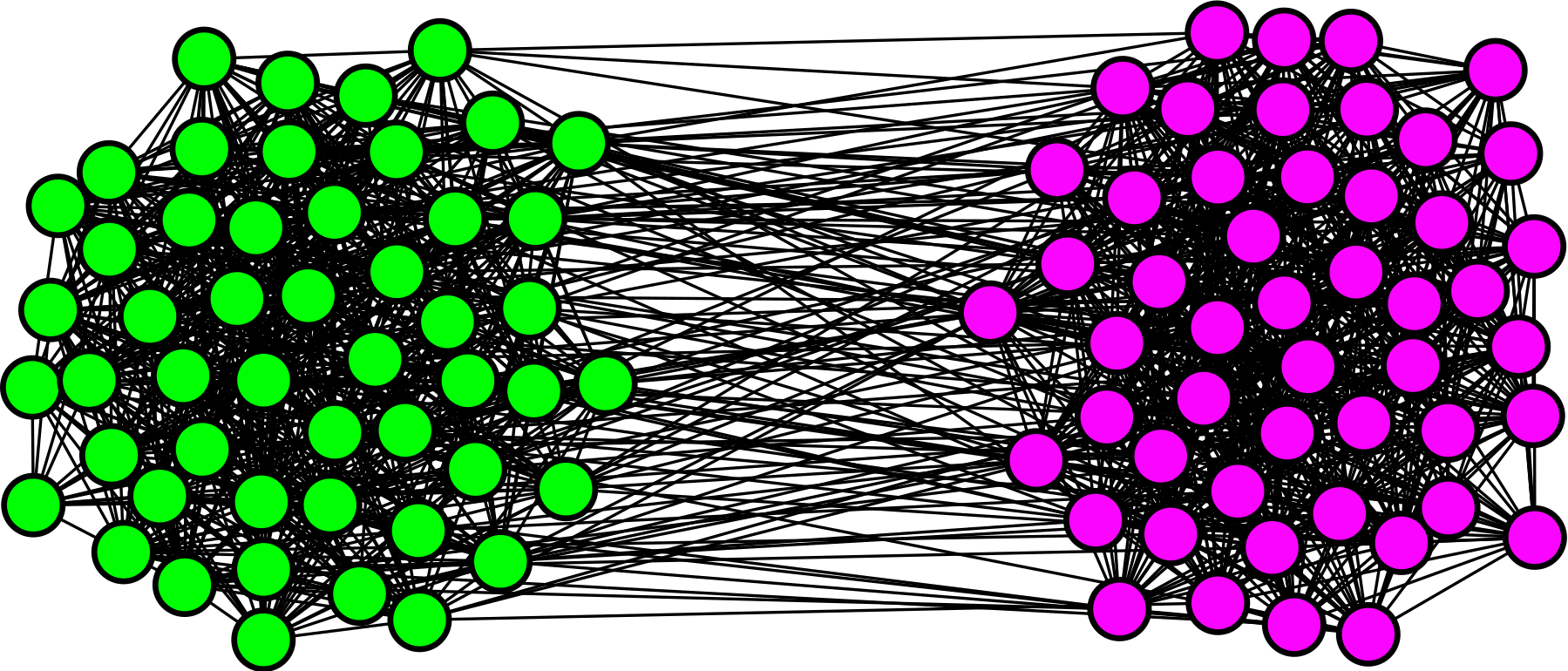}
    \caption{Positive edges}
\end{subfigure}
\qquad
\begin{subfigure}[t]{0.45\textwidth}
\centering
    \includegraphics[width=0.9\textwidth]{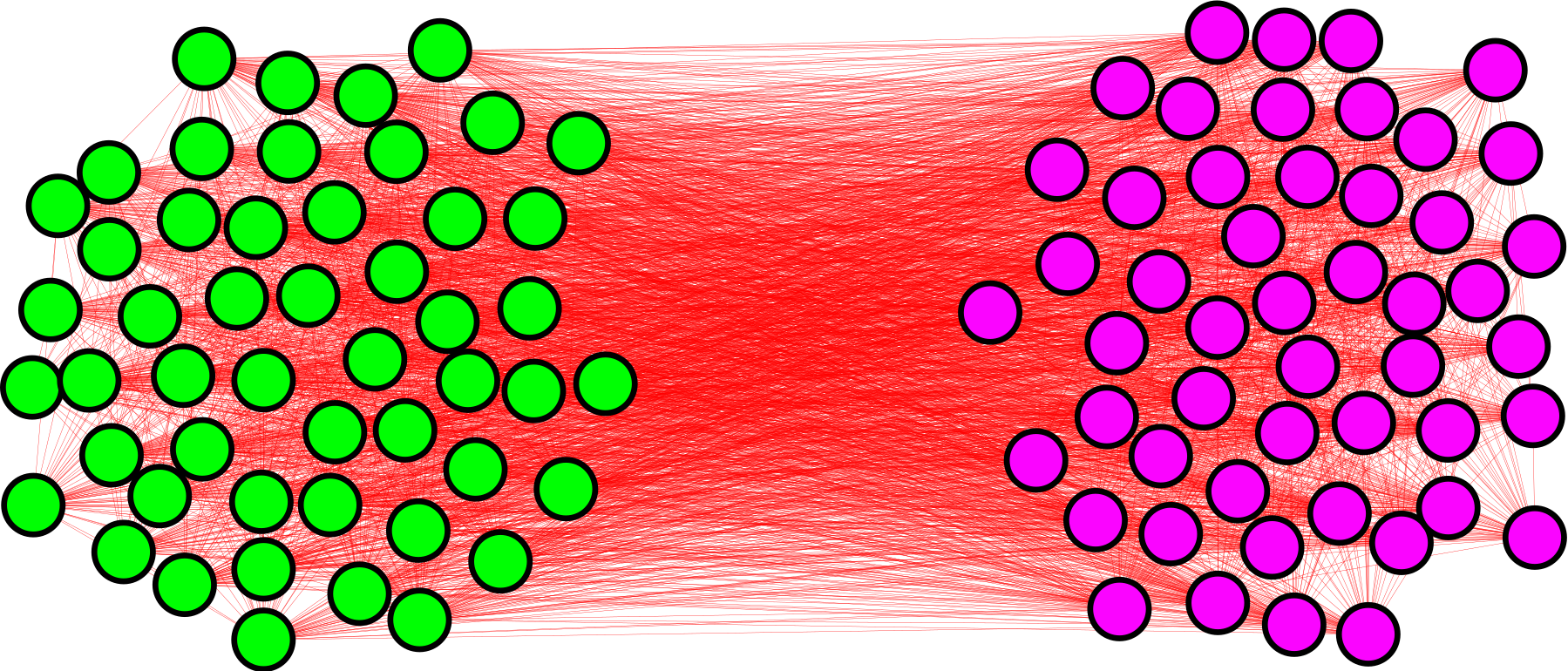}
    \caption{Negative edges}
\end{subfigure}
\caption{Exemplification of the topology emerging from $L^2$ on a realization of the Planted Partition with $100$ nodes and parameters $p_{in} = 0.5$ and $p_{out} = 0.05$. The positive edges coincide with the original structure but are reweighted. The negative ones appear between nodes that are initially at a 2-hop distance. It can be seen that a considerable amount of negative edges appear between clusters, bringing the potential to boost their detection.}
\label{L2_topology.fig}
\end{figure}
In this work, we propose to change the graph topology in which the problem is solved as a means to improve classification. We evoke such change by considering powers of the Laplacian matrix, noting that the $L^{\gamma}$ operator, for $\gamma > 0$, generates a new graph for every fixed $\gamma$ value. More precisely, the Laplacian definition indicates that $L^{\gamma} = V \Lambda^{\gamma} V^T = D_{\gamma} - W_{\gamma}$ codes for a new graph, where $\left[ D_{\gamma} \right]_{uu} = \left[L^{\gamma} \right]_{uu}$ refers to a generalized degree matrix and $\left[W_{\gamma}\right]_{uv} = - \left[ L^{\gamma} \right]_{uv} $, with $u \neq v$, to a generalized adjacency matrix that satisfies the Laplacian property $\left[D_{\gamma}\right]_{uu} = \sum_{v} \left[W_\gamma \right]_{uv}$ since $L^\gamma \mathbbm{1} = 0$. We refer to such graphs as $L^\gamma$-graphs.

The $L^\gamma$-graphs reweight the edges of the original structure and creates links between originally far-distant nodes. Indeed, for $\gamma \in \mathbb{Z}$ the new edges can be related to paths of different lengths. To have a grasp on this, let us take the topology from $\gamma = 2$ as an example: for $L^2 = (D - W)^2 = D^2+W^2-(DW+WD)$, the elements of the emanating graph are given as $[D_2]_{uu} = [L^2]_{uu} = D_{uu}^2 +\sum_{v} W_{uv}^2$ and $\left[W_{2}\right]_{uv} = - [L^2]_{uv} = (D_{uu}+D_{vv})W_{uv} - W^2_{uv} - \sum_{l\neq u,v} W_{ul}W_{lv}$, showing that, in $W_{2}$, nodes originally connected get their link reweighted (still, remaining positive) while those at a 2-hop distance become linked by a negatively weighted edge. 

This change in the topology has the potential to impact clustering, as the emergence of positive and negative edges opens the door for an interpretation in terms of an agreement (positive edge) or a disagreement (negative edge) between datapoints. Hence, clustering can be revamped to assume that nodes agreeing should belong to the same cluster and nodes disagreeing should belong to different ones. From this perspective, a revisit to the case of $\gamma = 2$ shows that this is indeed a potentially good topology since, for several graphs, it is more likely that vertices having a 2-hop distance lie in different clusters than in the same one, thus creating a considerable amount of disagreements between clusters, that may enhance their separability. This idea is illustrated in Figure \ref{L2_topology.fig}, where for a realization of the planted partition model we show that with $\gamma = 2$ a big amount of negative edges appear between clusters. 

Thus, in the reminder of the paper we investigate if for a target set of nodes $S_{gt}$, the detection of $S_{gt}$ can be enhanced by solving the clustering problem in some of these new graphs. 

\begin{remark}
 \normalfont The graphs emerging in the regime $0 < \gamma < 1$ have already been studied in \cite{Riascos.2014.fractional, DeNigris.2017.fractional, Bautista.2017.Levy}, where it is shown that such graphs remain within the class of graphs with only positive edges, hence preserving the random walk framework. In these works, the graphs were shown to embed the so-called L\'evy flight, permitting random walkers to perform long-distant jumps in a single step.
\end{remark}

\subsection{The $L^\gamma$-PageRank method}\label{LgPR-method.sec}
The signed graphs emerging from $L^\gamma$ preclude the employment of the random walk-based approaches to find clusters, as `negative transitions' appear. Thus, the $L^\gamma$-graphs call for a technique to find clusters in such graphs. In this subsection, we introduce $L^\gamma$-PageRank, a generalization of PageRank to finds clusters on the $L^\gamma$-graphs. Further, we analyze the $L^\gamma$-PageRank theoretical properties and clustering capabilities.

For our analysis, it is useful to first extend some of the graph topological definitions to the $L^\gamma$-graphs. Let $vol_{\gamma}(S) = \sum_{u \in S} \left[D_\gamma \right]_{uu}$ denote the generalized volume of $S$. Let $\pi_{\gamma}$ denote a generalized stationary distribution with entries given by $(\pi_{\gamma})_u = \left[D_{\gamma}\right]_{uu} / vol_{\gamma}(G)$. It is important to stress that $[D_{\gamma}]_{uu} = \sum_k \lambda_k^\gamma Q_{uk}^2 \geq 0$. Thus, for all $\gamma > 0$, the generalized volume and the generalized stationary distribution are non-negative quantities. 

The Cheeger ratio metric, lacking the ability to account for the sign of edges, cannot be employed to assess the presence of clusters in the $L^\gamma$-graphs. Thus, we generalize the Cheeger ratio definition to the new graphs as follows. 
\begin{definition}
For a set of nodes $S \subseteq V$, the generalized Cheeger ratio, or generalized conductance, of $S$ is defined as
\begin{equation}
	h_S^{(\gamma)} = \frac{ \sum_{u \in S} \sum_{v \in S^c} \left[W_{\gamma}\right]_{uv} }{ \min \left( vol_{\gamma}(S), vol_{\gamma}(S^c) \right) }
\end{equation}
\end{definition}
This generalization of the Cheeger ratio is mathematically sound. First, it is a non-negative quantity since $\sum_{u \in S} \sum_{v \in S^c} \left[W_{\gamma}\right]_{uv} = \mathbbm{1}_S^T L^\gamma \mathbbm{1}_S \geq 0$. Second, the set $S$ attaining the minimum value coincides with a sensible clustering. To show the latter, let the edges in $W_{\gamma}$ be split according to their sign as ${W_\gamma} = {W_\gamma}^{+} + {W_\gamma}^{-}$. Let $\mathcal{A}_{in}(S) = \sum_{u \in S}\sum_{w \in S} |\left[W_\gamma^+\right]_{uw}|$ be the sum of agreements within $S$, $\mathcal{A}_{out}(S) = \sum_{u \in S}\sum_{v \in S^c} |\left[W_\gamma^+\right]_{uv}|$ the agreements between $S$ and $S^c$, $\mathcal{D}_{in}(S) = \sum_{u \in S}\sum_{w \in S} |\left[ W_\gamma^-\right]_{uw}|$ the disagreements within $S$, and $\mathcal{D}_{out}(S) = \sum_{u \in S}\sum_{v \in S^c} |\left[ W_\gamma^- \right]_{uv}|$ the disagreements between $S$ and $S^c$. Then we state the following lemma.
\begin{lemma}\label{GCR-Interpretation.lemma}
Let $S^* = \argmin_S h_S^{(\gamma)}$. Then, $S^*$ also maximizes $\mathcal{D}_{out}(S^*)$ and $\mathcal{A}_{in}(S^*)$.
\end{lemma}
\noindent The proof is provided in Appendix \ref{GCR-Interpretation.proof}.  

Lemma \ref{GCR-Interpretation.lemma} shows that, for clustering in the $L^\gamma$-graphs, it is good to search for sets with small generalized Cheeger ratio as those sets have strong between-cluster disagreements and strong within-cluster agreements.

Now, we introduce the $L^\gamma$-PageRank formulation. Departing from the optimization problem in (\ref{PR-optimization.eq}), we revamp PageRank to operate on the $L^\gamma$-topology as follows.
\begin{definition}
The $L^{\gamma}$-PageRank G-SSL is defined as the solution to the optimization problem:
\begin{equation}\label{LgPR-optimization.eq}
\argmin_{f} \left\{ f^T D_{\gamma}^{-1} L^{\gamma} D_{\gamma}^{-1} f + \mu (f - y)^T D_{\gamma}^{-1} (f - y) \right\}
\end{equation}
\end{definition}
The two following Lemmas show that, for any $\gamma > 0$, the $L^\gamma$-PageRank solution exists in closed form and such solution preserves the PageRank properties. 
\begin{lemma}\label{LgPR-solution.lemma}
Let $\gamma > 0$. Then, problem (\ref{LgPR-optimization.eq}) is convex with closed form solution given by
\begin{equation}\label{LgPR-solution.eq}
f = \mu \left( L^{\gamma}D_{\gamma}^{-1} + \mu \mathbb{I} \right)^{-1} y 
\end{equation}
\end{lemma}
\noindent The proof is provided in Appendix \ref{LgPR-solution.proof}.
\begin{remark}
 \normalfont Eq. (\ref{LgPR-solution.eq}) emphasizes the difference between our approach and the one in \cite{Zhou.2011.semi}: they propose to iterate the operator in the G-SSL solution as $f = \mu \left( \left[ LD^{-1} \right]^m+ \mu \mathbb{I} \right)^{-1} y$, for $m \in \mathbb{Z}_{> 0}$, for which the formulation of the optimization problem having this expression as solution remains unknown. 
\end{remark}
\begin{remark}
\normalfont The solution of $L^\gamma$-PageRank in Eq. (\ref{LgPR-solution.eq}) can be easily cast as a low-pass graph filter, allowing a fast and distributed approximation via Chebyshev polynomials \cite{Shuman.2018.distributed}.
\end{remark}

\begin{lemma}\label{LgPR-properties.lemma}
Let $\gamma > 0$. The $L^\gamma$-PageRank solution in (\ref{LgPR-solution.eq}) satisfies the following properties: (\textit{i}) mass preservation: $\sum_{u \in \mathcal{V}} f_u  = \sum_{u \in \mathcal{V}} y_u$; (\textit{ii}) stationarity: $f = \pi_{\gamma}$ if $y = \pi_{\gamma}$; and (\textit{iii}) limit behavior: $f \to \pi_{\gamma}$ as $\mu \to 0$ and $f \to y$ as $\mu \to \infty$. 
\end{lemma}
\noindent The proof is provided in Appendix \ref{LgPR-properties.proof}.

The previous Lemmas are important because they show that our generalization, for any $\gamma > 0$, is a well-posed problem. Indeed, the properties of Lemma \ref{LgPR-properties.lemma} imply that, while not necessarily modeled by random walkers, $L^{\gamma}$-PageRank remains a diffusion process having $\pi_{\gamma}$ as stationary state and diffusion rate controlled by the $\mu$ parameter.

Our next results shows that it is hard for such diffusion process to escape clusters in the $L^\gamma$-graphs.
\begin{lemma}\label{LgPR-bound.lemma}
Let $\gamma > 0$ and let $S \subset \mathcal{V}$ be an arbitrary set with $vol_{\gamma}(S) \leq vol_{\gamma}(G)/2$. For a labeled point placed at node $u \in S$ with probability proportional to its generalized degree in $S$, i.e. $\frac{[D_{\gamma}]_{uu}}{vol_{\gamma}(S)}$, $L^\gamma$-PageRank satisfies
\begin{equation} 
	\mathbb{E}\left[f(S^c) \right] \leq \frac{h_S^{(\gamma)}}{\mu} 
\end{equation} 
\end{lemma} 
\noindent The proof is provided in Appendix \ref{LgPR-bound.proof}.

Lemma \ref{LgPR-bound.lemma} admits a similar interpretation as Lemma \ref{PR-bound.lemma}. Namely, if $L^\gamma$-PageRank is applied to the labeled points of some set $S$ with small $h_S^{(\gamma)}$, then diffusion is confined to $S$ and the score values outside of $S$ are expected to be small. Thus, by looking at the nodes with largest score values we should be able to retrieve a good estimation of $S$. If such score concentration phenomenon takes place, then a sharp drop must appear after sorting the $L^\gamma$-PageRank scores in descending order. We will use the following lemma to show that if a sharp drop is present, then the sweep cut procedure applied on the $L^\gamma$-PageRank vector retrieves a partition $\hat{S}$ that has small $h_{\hat{S}}^{(\gamma)}$.
\begin{lemma}\label{LgPR-sharpdrop-ineq.lemma}
Let $q$ denote the permutation vector and $S_j$ denote the set associated to $q_j$ obtained by applying the sweep-cut procedure on the $L^\gamma$-PageRank vector. Then, the partition $\mathcal{V} = S_j \cup S_j^c$ satisfies the inequality:
\begin{flalign}
&\mathcal{A}_{out}(S_j) \left(2 - \frac{(q_j - q_{j+1})}{(q_1 - q_N)}\right) -  \mathcal{D}_{out}(S_j)\left(2\frac{(q_j - q_{j+1})}{(q_1 - q_N)} - 1\right) \geq    \frac{\mu \left( y(S_j) - f(S_j) \right)}{(q_1 - q_n)} \notag\\
&\qquad \geq \mathcal{A}_{out}(S_j) \left(2\frac{(q_j - q_{j+1})}{(q_1 - q_N)} - 1\right) -  \mathcal{D}_{out}(S_j)\left(2 - \frac{(q_j - q_{j+1})}{(q_1 - q_N)}\right) &&
\end{flalign}
\end{lemma}
\noindent The proof is provided in Appendix \ref{LgPR-sharpdrop-ineq.proof}.

We have that $\sum_{u \in S_j} \sum_{v \in S_j^c} \left[W_{\gamma}\right]_{uv} = \mathcal{A}_{out}(S_j) - \mathcal{D}_{out}(S_j) \geq 0$. Thus, the generalized Cheeger ratio of $S_j$ is small if $\mathcal{A}_{out}(S_j)$ is not much larger than $\mathcal{D}_{out}(S_j)$. In the inequality above, we have two cases in which $(q_j - q_{j+1})/(q_1 - q_N) \approx 1$: (a) $q$ is approximately constant; and (b) $q$ has a drop that satisfies $q_j \approx q_1$ and $q_{j+1} \approx q_N$. The former can only occur if $f \to \pi_\gamma$ and clearly no cluster can be retrieved from that vector, as confirmed by the inequality growing unbounded. The latter case is what we coin as having a sharp drop between $q_j$ and $q_{j+1}$. In such case, the inequality is controlled by the difference $y(S_j) - f(S_j)$ which, due to the mass preserving property and the assumption that $q_{j+1} \approx q_N$, should be small. Thus, granting that $\mathcal{A}_{out}(S_j)$ is not much larger than $\mathcal{D}_{out}(S_j)$ and $S_j$ has a small $h_{S_j}^{(\gamma)}$.

{\bf Discussion.}~The previous results show that $L^\gamma$-PageRank is a sensible tool to find clusters in the $L^\gamma$-graphs, i.e. groups of nodes with small generalized Cheeger ratio. Thus, revisiting the classification case in which we target group of nodes $S_{gt}$, we have that the smaller the value of $h_{S_{gt}}^{(\gamma)}$, the better the $L^\gamma$-PageRank method can recover it. This observation, in addition to noting that standard PageRank emerges as the particular case of $\gamma = 1$, indicate that we should be able to enhance the performance of G-SSL in the detection of $S_{gt}$ by finding the graph, i.e. the $\gamma$ value, in which $h_{S_{gt}}^{(\gamma)} < h_{S_{gt}}^{(1)}$.

\subsection{The selection of $\gamma$}\label{g-tunig.sec}
\subsubsection{Case of $\gamma = 2$: analytic study}
In Section \ref{Lg-topology.sec}, it was argued that the topology emerging from $L^2$ places a negatively weighted link between nodes at a 2-hop distance, thus carrying the potential to place a big amount of disagreements between clusters that may enhance their separability. Our next result formalizes this claim, demonstrating that on graphs from the Planted Partition model it is expected that the $L^2$-graph improves the generalized Cheeger ratio. 
\begin{theorem}\label{SBM-L2.theorem}
	Consider a Planted Partition model of parameters ($p_{in}$, $p_{out}$) and cluster sizes $|S_{gt}| = |S_{gt}^c| = n$. Then, as $n \to \infty$ we have that 
	\begin{equation} 
		\mathbb{E} \left[h^{(2)}_{S_{gt}}\right] = 2 \mathbb{E}\left[ h_{S_{gt}}^{(1)} \right]^2, 
	\end{equation} 
	where $\mathbb{E}\left[h_{S_{gt}}^{(1)} \right] = p_{out}/(p_{in} + p_{out})$. 
\end{theorem}
\noindent The proof is provided in Appendix \ref{SBM-L2.theorem.proof}.
\begin{corollary}\label{SBM-L2.corollary}
	 If $p_{in} \geq p_{out}$, then $\mathbb{E} \left[h^{(2)}_{S_{gt}}\right] \leq \mathbb{E}\left[ h_{S_{gt}}^{(1)} \right]$, with equality occurring in the case $p_{in} = p_{out}$.  
\end{corollary}
\noindent The proof is provided in Appendix \ref{SBM-L2.corollary.proof}. 

Theorem \ref{SBM-L2.theorem} and Corollary \ref{SBM-L2.theorem} open the door to investigate, on arbitrary graphs, in which cases the $L^2$-graph improves the generalized Cheeger ratio of a set. In the next Proposition, we provide a sufficient condition in which the $L^2$-graph improves the generalized Cheeger ratio a set.
\begin{proposition}\label{Condition-L2.proposition}
	Let $\langle D_{S_{gt}} \rangle$ denote the mean degree of $S_{gt}$. A sufficient condition on $S_{gt}$ so that $h^{(2)}_{S_{gt}} \leq h_{S_{gt}}^{(1)}$ is 
	\begin{equation}
		\langle D_{S_{gt}} \rangle \geq \max_{u \in S_{gt}} \sum_{v \in S_{gt}^c} W_{uv} + \max_{w \in S_{gt}^c} \sum_{\ell \in S_{gt}} W_{w \ell},
	\end{equation}
\end{proposition}
\noindent The proof is provided in Appendix \ref{Condition-L2.proof}.

This proposition points in the same direction as Theorem \ref{SBM-L2.theorem}, saying that graphs having a cluster structure are bound to benefit from $L^2$. Concretely, the first term on the right hand side of the inequality searches, among all the nodes of $S_{gt}$, the one that has the maximum number of connections towards $S_{gt}^c$. The second term does the reverse for the nodes of $S_{gt}^c$. Hence, asking for the nodes of $S_{gt}$ to have, on average, more connections than the maximum possible boundary implies that $S_{gt}$ should have a cluster structure.

\subsubsection{An algorithm for the estimation of the optimal $\gamma$}
\begin{figure}[t]
    \centering
    \begin{subfigure}[t]{0.31\textwidth}
    \includegraphics[width=1\textwidth]{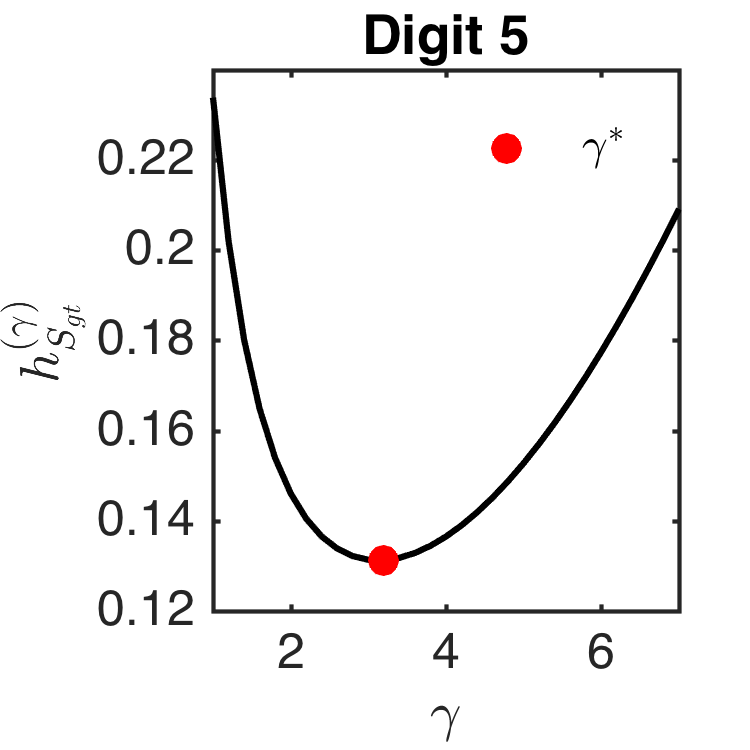}
    \caption{$\gamma^* = \argmin_\gamma h_{S_{gt}}^{(\gamma)}$}
    \label{MNIST_CheegerRatio-1.fig}
    \end{subfigure}
    \begin{subfigure}[t]{0.31\textwidth}
    \includegraphics[width=1\textwidth]{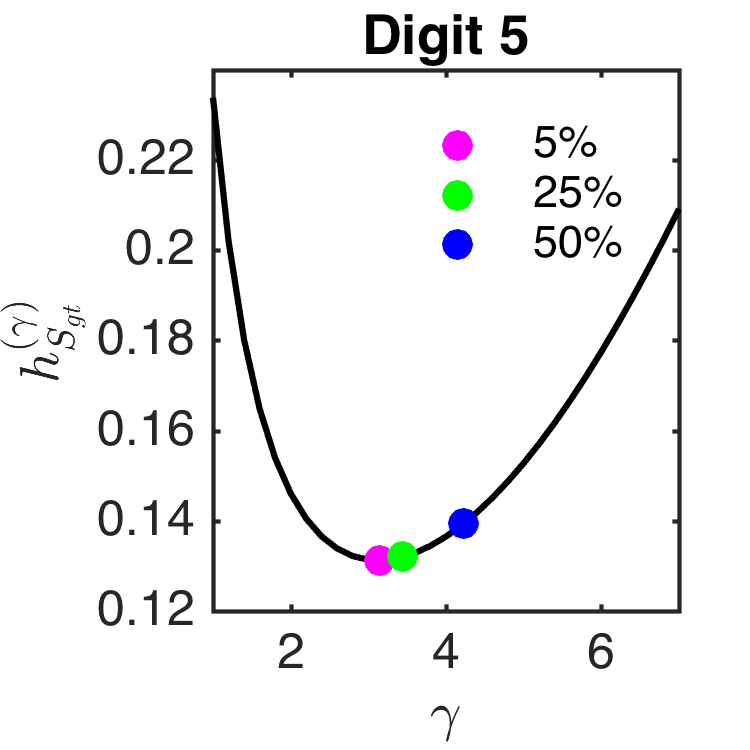}
    \caption{$\gamma^*$ on subsets of $S_{gt}$}
    \label{MNIST_CheegerRatio-2.fig}
    \end{subfigure}
    \begin{subfigure}[t]{0.31\textwidth}
    \includegraphics[width=1\textwidth]{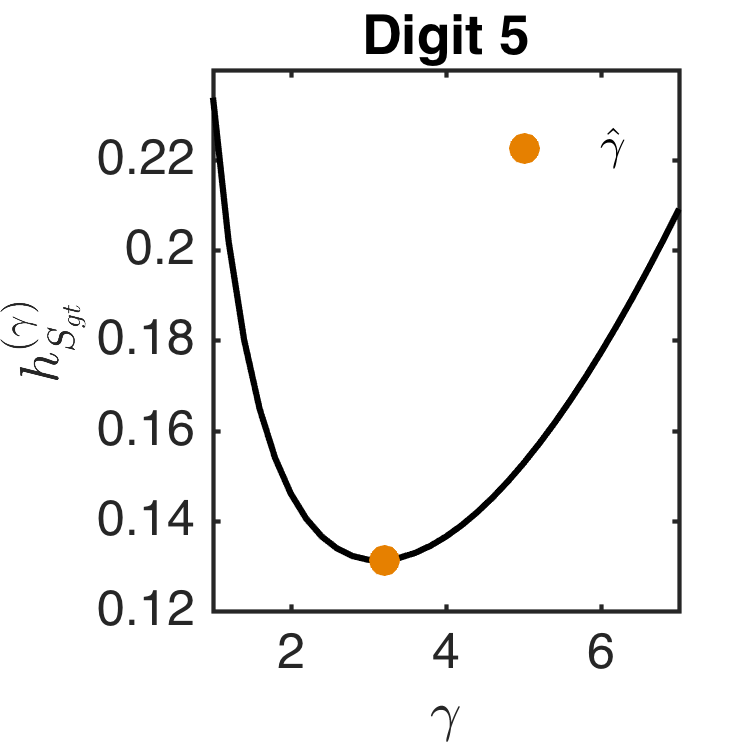}
    \caption{$\hat{\gamma}$ by algorithm}
    \label{MNIST_CheegerRatio-3.fig}
    \end{subfigure}
    \caption{Generalized Cheeger ratio of $S_{gt}$ as a function of $\gamma$. For the plot, $S_{gt}$ is a digit of the MNIST dataset.}
    \label{MNIST_CheegerRatio.fig}
\end{figure}
\noindent Numerical experiments show that increasing $\gamma$ can further decrease the generalized Cheeger ratio up to a point where it starts increasing. We show an example of this phenomenon in Figure \ref{MNIST_CheegerRatio-1.fig}, displaying the evolution of $h_{S_{gt}}^{(\gamma)}$ as a function of $\gamma$ when $S_{gt}$ corresponds to a digit of the MNIST dataset. From the figure, it is evident that an optimal value appears, denoted $\gamma^* = \argmin_\gamma h_{S_{gt}}^{(\gamma)}$, raising the question of how to find such value. Since the behavior of $h_{S_{gt}}^{(\gamma)}$ depends on $S_{gt}$, in practice, the derivative or a greedy search to find $\gamma^*$ cannot be employed since $S_{gt}$ is unknown. A second question that arises is whether the optimal value changes drastically or smoothly with changes in $S_{gt}$. We perform the following test: for a given $S_{gt}$ (same MNIST digit), we remove some percentage of the nodes in $S_{gt}$ and record the optimal value on subsets of $S_{gt}$. More precisely, recall that $h_{S_{gt}}^{(\gamma)} = \mathbbm{1}_{S_{gt}}^T L^\gamma \mathbbm{1}_{S_{gt}} / \mathbbm{1}_{S_{gt}}^T D_\gamma \mathbbm{1}_{S_{gt}}$, hence we randomly select some percentage of the entries indexing $S_{gt}$ in $\mathbbm{1}_{S_{gt}}$, set them to zero and obtain a new indicator function indexing a subset of $S_{gt}$. Mean results are evaluated in the original curve and displayed in Figure \ref{MNIST_CheegerRatio-2.fig}. The figure suggest that it is not necessary to know $S_{gt}$ to find a proxy $\hat{\gamma}$ of $\gamma^*$, it suffices to know a subset of $S_{gt}$. Based on the last observation, we propose Algorithm \ref{estimation.alg} for the estimation of $\gamma^*$. The rationale of the algorithm is to exploit the labeled points and the graph to find a proxy $\hat{S}$ of $S_{gt}$ on which we can compute the estimate. The procedure consists in letting walkers started from the label points, run for a number of steps that is determined by the maximum geodesic distance between the labels. This allows walkers to explore $S_{gt}$ without escaping too far from it. After running the walk, we list the nodes in descending order according to the probability of finding a walk at a node. We take the first element on the list (the one where it is more likely to find a walker), add it to $\hat{S}$ and remove it from the list, so that the former second element becomes the first in the listing. We repeat the procedure until the probability of finding a walker in the nodes conforming $\hat{S}$ is 0.7. 
\begin{algorithm}[t]
\caption{Estimation of $\gamma^*$}
\label{estimation.alg}
\begin{algorithmic}
        \State \textbf{Input}: $\mathcal{G}, \mathcal{V}_{S_{gt}}$ and a grid of $\gamma$ values.
        \State \textbf{Output}: $\hat{\gamma}$
        \State Compute $\Delta_{uv} ~ \forall ~ u, v \in \mathcal{V}_{S_{gt}}$.
        \State Set $k = \max_{u,v} \left\{\Delta_{uv}\right\}$
        \State Set $\chi = y / \|y\|_1$
        \State Run a $k$-step walk with seed $\chi$: $x^T = \chi^T P^k$
        \State Reorder the vertices as $v_{1}, \cdots, v_N$, so that $x_{v_i} \geq x_{v_{i+1}}$
        \For{$i = 1 : N$}
            \If{$\sum_{j = 1}^{i} x_{v_j} < 0.7$}
                \State Set $(\mathbbm{1}_{\hat{S}})_{v_i} = 1$ 
            \Else
                \State Set $(\mathbbm{1}_{\hat{S}})_{v_i} = 0$ 
            \EndIf
        \EndFor
        \State Compute $h_{\hat{S}}^{(\gamma)} = \frac{ \mathbbm{1}_{\hat{S}}^T L^\gamma \mathbbm{1}_{\hat{S}}} {\mathbbm{1}_{\hat{S}}^T D_{\gamma} \mathbbm{1}_{\hat{S}} } ~ \forall ~ \gamma$
        \State Return $\hat{\gamma} =  \argmin_{\gamma} h_{\hat{S}}^{(\gamma)}$.
\end{algorithmic}
\end{algorithm}

In Table \ref{alg_evaluation.table}, we evaluate the performance of Algorithm \ref{estimation.alg} on the estimation of $\gamma^*$ for all the digits of the MNIST. The first row displays, as $\gamma^*$, the value of $\gamma$ (from the input range) attaining the minimum generalized Cheeger ratio. The second row displays the performance of the algorithm when estimating such value. The last three rows show the value of the generalized Cheeger ratio evaluated at $\gamma^*$, $\hat{\gamma}$ and $\gamma = 1$, respectively. The estimator finds values of $\hat{\gamma}$ whose Cheeger ratios are: (a) significantly smaller than those of $\gamma = 1$; (b) close to the optimal. 

\begin{table}[t]
  \centering
    \begin{tabular}{M{7mm} *{9}{M{8mm}}} \toprule
{\bf{Digit}}                    & {\bf{1}}     & {\bf{2}}     & {\bf{3}}     & {\bf{4}}     & {\bf{5}}     & {\bf{6}}    &  {\bf{7}}    & {\bf{8}}   & {\bf{9}}\\ \midrule 
$\gamma^*$                      & 7.0   & 3.0   & 7.0   & 3.2   & 3.2   & 7.0     & 7.0   & 3.2  & 4.2 \\\cmidrule(l){2-10} 
$\hat{\gamma}$                  & 5.45 (0.15) & 3.10 (0.14) & 6.41 (0.11) & 4.92 (0.16) & 3.20 (0.14) & 6.04 (0.15) & 4.98 (0.17) &  4.40 (0.18) & 5.08 (0.15) \\\cmidrule(l){2-10} 
$h_{S_{gt}}^{(\gamma^*)}$       & 0.065 & 0.166 & 0.035 & 0.141 & 0.131 & 0.011 & 0.052 & 0.116 & 0.135 \\\cmidrule(l){2-10} 
$h_{S_{gt}}^{(\hat{\gamma})}$   & 0.073 (9e-4) & 0.174 (8e-4) & 0.041 (1e-3) & 0.185 (4e-3) & 0.148 (2e-3) & 0.017 (1e-3) & 0.074 (2e-3) & 0.142 (2e-3) & 0.149 (9e-4) \\\cmidrule(l){2-10}  
$h_{S_{gt}}^{(1)}$              & 0.175 & 0.248 & 0.216 & 0.258 & 0.233 & 0.107 & 0.203 & 0.215 & 0.285 \\ \bottomrule
    \end{tabular}
\caption{Evaluation of Algorithm \ref{estimation.alg} on the MNIST Dataset. Mean values (95\% confidence interval) are shown. The graph construction guidelines are provided in Section \ref{Performance_RealData.sec}. For the experiment, 500 realizations of labeled points and a grid of $\gamma$ ranging from 1 to 7 with a resolution of 0.2 were used.}
\label{alg_evaluation.table}
\end{table}

\section{$L^\gamma$-PageRank in practice}\label{Performance.sec}
\subsection{Planted Partition}\label{Performance_PlantedPartition.sec}
{\bf Experimental setup and goals.}~In the following experiment, we show that $L^\gamma$-PageRank can increase the performance of G-SSL as the graph approaches the Planted Partition detectability transition. More precisely, it is shown in \cite{Mossel.2015.reconstruction} that the Planted Partition possesses a detectability threshold above which unsupervised methods are unable to retrieve a meaningful clustering. Indeed, if the clusters sizes are denoted as $|S_{gt}| = |S_{gt}^c| = n$, the mean degree of a node is given as $C_{avg} = C_{in} + C_{out}$, where $C_{out} = (p_{out})(n)$ and $C_{in} = (p_{in})(n-1)$. It is then possible to recover a cluster that is positively correlated with the true partition, in an unsupervised manner, if $(C_{in} - C_{out})^2 > 2 (C_{in} + C_{out})$, and impossible otherwise. As for G-SSL, the work in \cite{Zhang.2014.phase} showed that such threshold can be overcome when a fraction of labeled points is introduced to the task. Nonetheless, the performance of G-SSL drastically degrades when approaching the detectability transition.

The experimental setup is the following: for a given $C_{out}/C_{in}$, a realization of the Planted Partition is drawn with $n = 500$ and $C_{avg} = 3$. Then, 1\% of labeled points are sampled at random and the $L^\gamma$-PageRank method is applied for different values of $\mu$ lying on a discrete grid. The clusters are determined via a sweep-cut procedure, and the best performance is retained. The whole procedure is repeated for 10 different realizations of the labeled points. Finally, all the preceding steps are repeated for 100 graph realizations. Performance is assessed in terms of the Matthews Correlation Coefficient (MCC) \cite{Matthews.1975.comparison}, so that a value of 1 implies perfect agreement with the true partition and 0 a random decision. 

\begin{figure}[t]
    \centering
    \begin{subfigure}[t]{0.45\textwidth}
    \includegraphics[width=1\textwidth]{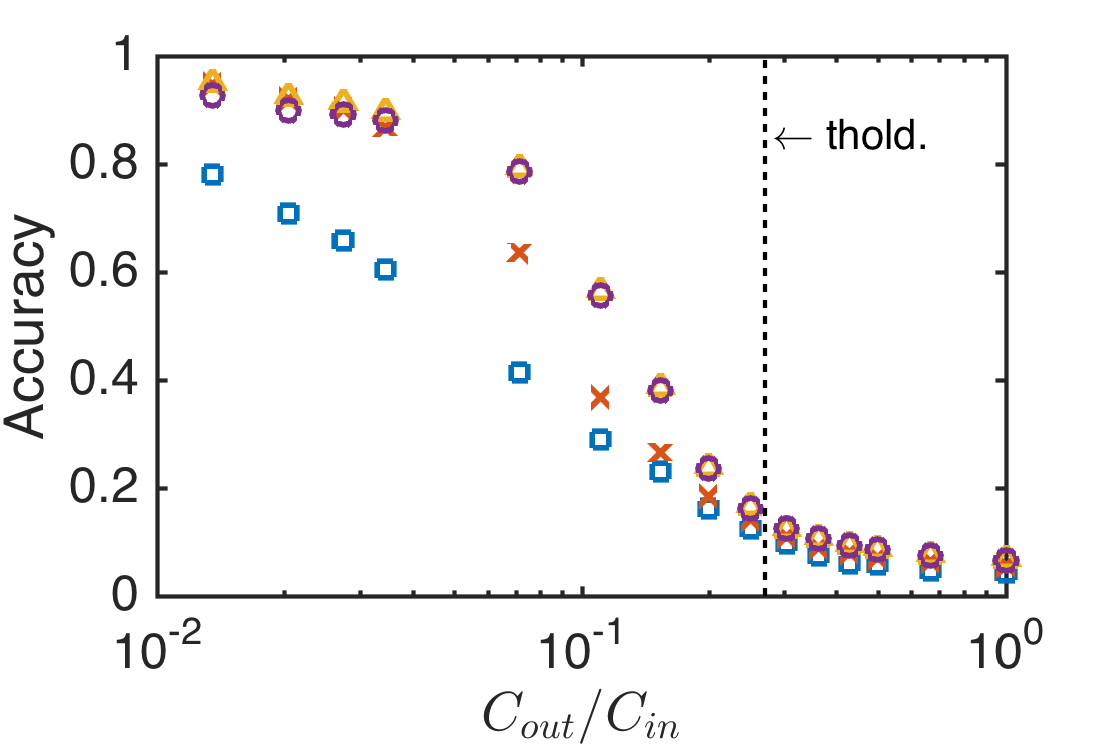}
    \end{subfigure}
    \begin{subfigure}[t]{0.45\textwidth}
    \includegraphics[width=1\textwidth]{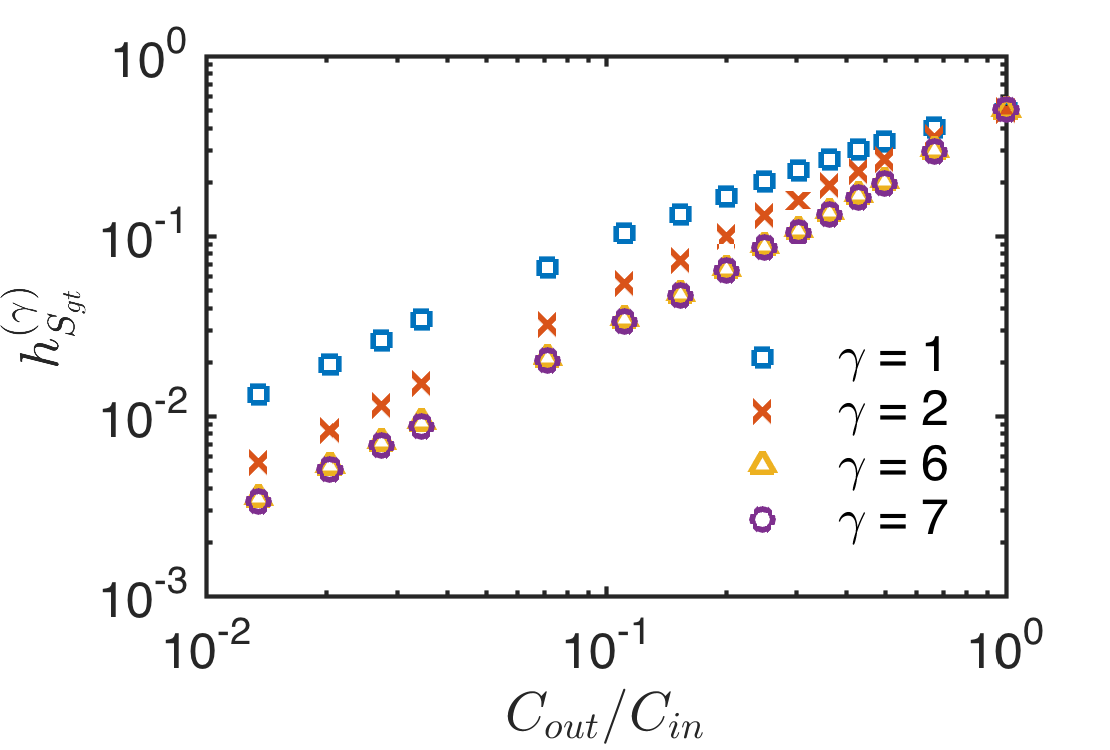}
    \end{subfigure}
    \caption{Improved detection of the Planted Partition.}
    \label{Perforance_PlantedPartition.fig}
\end{figure}

\noindent{\bf Results and discussion.}~Figure \ref{Perforance_PlantedPartition.fig} displays the performance of $L^\gamma$-PageRank at recovering the Planted Partition as a function of the ratio $C_{out}/C_{in}$. Standard PageRank ($\gamma = 1$) performs poorly as the configuration approaches the phase transition (referred by the vertical line) since $h_S^{(1)}$ becomes large. Clearly, the introduction of $\gamma$ allows to decrease $h_{S_{gt}}^{(\gamma)}$, which, accordingly, enhances the clustering performance. Furthermore, the figure verifies that the smaller the value of $h_{S_{gt}}^{(\gamma)}$ (right plot), the better the $L^\gamma$-PageRank recovers the true partition (left plot). It is important to remark that, for this experiment, while $\gamma = 2$ shows good improvements, larger values of $\gamma$ keep improving $h_S^{(\gamma)}$, until it reaches a saturation plateau, designating a region of optimal $\gamma$ values ($\gamma \geq 6$). 

\subsection{Real world datasets}\label{Performance_RealData.sec}
\afterpage{
\begin{table}[t]
  \centering
    \begin{tabular}{p{10mm}p{14mm}M{16mm}M{16mm}M{22mm}M{22mm}} \toprule
    & \qquad $S_{gt}$ \qquad &  $\gamma = 1$ & $\gamma = 2$ &  $\gamma = \hat{\gamma}$ & $\gamma = \gamma^*$ \\ \midrule 
                                    & Digit 1 & 0.67 (0.075) & 0.78 (0.032) & 0.78 (0.034) [5.4] & 0.80 (0.027) [7.0]  \\ \cmidrule(l){2-6}
                                    & Digit 2 & 0.38 (0.042) & 0.60 (0.064) & 0.64 (0.059) [3.3] & 0.64 (0.059) [3.0] \\ \cmidrule(l){2-6}
                                    & Digit 3 & 0.47 (0.040) & 0.61 (0.032)  & 0.61 (0.028) [6.0] & 0.61 (0.028) [7.0]\\ \cmidrule(l){2-6}
                                    & Digit 4 & 0.39 (0.022) & 0.48 (0.036) & 0.53 (0.044) [4.7] & 0.53 (0.037) [3.2] \\ \cmidrule(l){2-6}
     MNIST                          & Digit 5 & 0.44 (0.036)  & 0.56 (0.046) &                                   0.61 (0.036) [3.3] & 0.64 (0.035) [3.2] \\                                              \cmidrule(l){2-6}
                                    & Digit 6 & 0.90 (0.039)  & 0.94 (0.003) & 0.94 (0.002) [6.0] & 0.94 (0.002) [7.0] \\ \cmidrule(l){2-6}
                                    & Digit 7 & 0.43 (0.027)  & 0.66 (0.043) & 0.71 (0.042) [4.8] & 0.75 (0.032) [7.0] \\ \cmidrule(l){2-6}
                                    & Digit 8 & 0.47 (0.062)  & 0.65 (0.057) & 0.74 (0.038) [4.8] & 0.72 (0.050) [3.2] \\ \cmidrule(l){2-6}
                                    & Digit 9 & 0.43 (0.020)  & 0.52 (0.026) & 0.53 (0.023) [4.9] & 0.56 (0.026) [4.2] \\ \midrule
\multirow{2}{10mm}{Gender images}   & Female    & 0.51 (0.039) & 0.57 (0.028) & 0.57                                         (0.020) [3.0] & 0.57 (0.028) [2.0] \\  \cmidrule(l){2-6}
                                    & Male      & 0.55 (0.028) & 0.61 (0.021) & 0.60 (0.022) [3.3] & 0.61 (0.021) [2.4] \\ \midrule
\multirow{3}{10mm}{BBC articles}    & Business  &  0.80 (0.020) & 0.53 (0.038) & 0.72 (0.040) [1.3] &  0.81 (0.021) [1.1] \\ \cmidrule(l){2-6}
                                    & Entmt.   & 0.84 (0.027) & 0.57 (0.040) & 0.76 (0.047) [1.5] & 0.86 (0.025) [1.3]  \\ \midrule
\multirow{2}{10mm}{Phoneme}         & Nasal     & 0.37 (0.030) & 0.41 (0.028) & 0.43 (0.025) [2.9] & 0.43 (0.025) [3.0] \\     \cmidrule(l){2-6}
                                    & Oral      & 0.41 (0.025) & 0.44 (0.022) & 0.46 (0.019) [2.8] & 0.46 (0.019) [3.0] \\ \bottomrule
    \end{tabular}
    \caption{Performance on real world datasets: each cell reports MCC, 95\% confidence interval (parenthesis) and the value of $\gamma$ [squared brackets].}
    \label{Performance_RealData.table}
\end{table}}

{\bf Experimental setup and goals.}~In our following experiment, we assesses the performance of $L^\gamma$-PageRank and Algorithm \ref{estimation.alg} on real world datasets.

The experimental setup is as follows: graphs are build connecting the K-Nearest Neighbors (KNN) with distances computed via the Gaussian kernel, so that the weight between points $\mathbf{x}_u$ and $\mathbf{x}_v$ is given by $W_{uv} = \exp\{ - || \mathbf{x}_u - \mathbf{x}_v ||^2_2 /\sigma^2 \}$. For each class, 2\% of labeled points are randomly selected, $L^\gamma$-PageRank is applied for a grid of $\mu$ values, partitions are retrieved via the sweep-cut, and the best performance, assessed in terms of MCC, is retained. Such procedure is repeated for 100 realization of labeled points, except for the MNIST on which 30 realizations only are employed. In all cases, classes are balanced in size and the graph construction parameters are selected to provide a good distribution of weights as follows: (a) MNIST \cite{Lecun.1998.gradient}: Images of handwritten digits (1 to 9). From the entire dataset, 200 images of each digit are selected and used to build the graph with KNN = 10 and $\sigma = 10^4$; (b) Gender Images \cite{Hond.1997.distinctive}: Images of male and female subjects for gender recognition. From the entire dataset, 200 images of each gender are selected and used to build the graph with KNN = 60 and $\sigma = 10^4$. The large value of KNN is to avoid disconnected components; (c) BBC articles \cite{Greene.2006.practical}: Word frequency attributes from news media articles. From the entire dataset, 200 business and 200 entertainment articles are used to build the graph with KNN = 5 and $\sigma = 50$; and (d) Phoneme \cite{phoneme}: Five attributes to discern nasal sounds from oral sounds. From the entire dataset, 200 oral and 200 nasal sounds are used to build the graph with KNN = 10 and $\sigma = 2$.

\noindent{\bf Results and discussion.}~
Table \ref{Performance_RealData.table} shows the performance of $L^\gamma$-PageRank on the classification of these real world datasets. Clearly, the introduction of $\gamma$ can significantly improve performance and, in general, the estimation $\hat{\gamma}$ performs close to the optimal value $\gamma^*$. It can be seen that some datasets are more sensitive to $\gamma$ than others. For instance, in the BBC articles we observe that a small change in $\gamma$, going from $\gamma = 1$ to $\gamma^* = 1.1$, increases performance, and going further to $\hat{\gamma} = 1.3$ and $\gamma = 2$ significantly worsens the classification. On the other hand, the MNIST dataset is less sensitive to $\gamma$, obtaining similar performances with larger variations in $\gamma$. 

It is important to stress that, thus far, we have assumed possession of the proper tuning of the diffusion rate ($\mu$) that attains the best results. However, when working with real data, clusters may have intricate local structures, e.g. sub-clusters, that play an important role in the way information diffuses, and that can make more difficult the finding of the optimal diffusion rate $\mu$. As a result, two clusters may have equal Cheeger ratios but one of them being harder to find if its local structure is complex. Digit 8 poses an example of this phenomenon, where the mean performance for $\hat{\gamma}$ is slightly better than that of $\gamma^*$. This anomaly can be explained as an aftereffect of using a finite grid on $\mu$: for some realization of labeled points, the best performance for $\gamma^*$ falls in a region not covered by the grid.

\subsection{Unbalanced labeled data}\label{Performance_UnbalancedLabels.sec}
{\bf Experimental setup and goals.}~In our last experiment, we show that $L^\gamma$-PageRank, adapted to the multi-class setting described in Section \ref{soa_multiClass.sec}, can improve the performance of G-SSL in the presence of unbalanced labeled data.

The experimental setup is as follows: graphs with two balanced classes (in size) are built using the datasets from the preceding experiments. The parameters of the graphs' construction follow the guidelines provided in Section \ref{Performance_RealData.sec}. For the Planted Partition, the configuration is $n = 200$, $C_{avg} = 3$, $C_{out} = 0.1$. Then, unbalanced labeled points are drawn at random: 2\% from one class and 6\% from the other. Lastly, $L^\gamma$-PageRank, in the multi-class setting, is applied for a grid of $\mu$ values and the best performance, assessed by MCC, is recorded. For the planted partition, the procedure is repeated over 15 realizations of the labeled points and for 100 graph realizations. For the other datasets, 100 realizations of labeled points are employed. 

\noindent{\bf Results and discussion.}~
Table \ref{Performance_UnbalancedLabels.table} displays the performance $L^\gamma$-PageRank in the presence of unbalanced labeled data. It is important to stress that, in this framework, a unique value of $\gamma^*$ is used to retrieve all the clusters at the same time, precluding the notion of an optimal $\gamma$ as defined in Section \ref{g-tunig.sec}. However, one value of $\gamma$ seems to perform better, we denote it as $\gamma = $ Best. The results confirm that the introduction of $\gamma$ helps to improve the classification in the presence of the unbalanced labeled data.  

\begin{table}[t]
  \centering
    \begin{tabular}{p{13mm} *{6}{M{14mm}}} \toprule
                & Planted Partition & MNIST 4vs9 & MNIST 3vs8 & BBC articles & Gender images & Phoneme \\ \midrule 
$\gamma = 1$    & 0.81 (1.1e-2) & 0.51 (1.5e-2) & 0.70 (1.4e-2) &  0.66 (1.8e-2) & 0.63 (2.1e-2) & 0.44 (2.3e-2) \\\cmidrule(l){2-7} 
$\gamma = 2$    & 0.87 (8.7e-3) & 0.56 (1.5e-2) & 0.76 (1.2e-2) &  0.92 (5.0e-3) & 0.73 (1.6e-2) & 0.48 (1.4e-2) \\\cmidrule(l){2-7} 
$\gamma = $ Best& 0.90 (7.0e-3) [6] & 0.57 (1.5e-2) [3] & 0.78 (1.2e-2) [4] &  0.93 (1.5e-3) [3] & 0.75 (1.7e-2) [3] & 0.48 (1.4e-2) [1.9] \\ \bottomrule
    \end{tabular}
\caption{Performance on unbalanced labaled data: each cell reports MCC, 95\% confidence interval (parenthesis) and the value of $\gamma$ [squared brackets].} 
\label{Performance_UnbalancedLabels.table}
\end{table}

\section{Conclusion}\label{Conclusion.sec}
This work proposed $L^\gamma$-PageRank, an extension of PageRank based on (non necessary integer) powers of the (combinatorial) Laplacian matrix. Our analysis shows that the added degree of freedom offers more versatility than standard PageRank, providing the potential to address some of the limitations of G-SSL. Precisely, we showed that when clusters are obtained via the sweep-cut procedure, $L^\gamma$-PageRank can significantly outperform standard PageRank. Further, we showed that the multi-class approach also benefits from our proposition, as performance was enhanced in the presence of unbalanced labeled data. These improvements were possible due to the $L^\gamma$ ($\gamma > 0$) operator coding for graphs whose topology can reinforce the separability of clusters. The richness of such graphs comes from the sign of edges, allowing to code for similarities but also to emphasize dissemblance between individuals. Thus, while 2 nodes can only be disconnected on the initial graph, they can `repulse' themselves in these topologies. Notably, we have shown that there is an optimal graph (related to an optimal $\gamma$) on which the classification will lead to a maximal performance. We proposed a simple yet efficient algorithm to estimate the optimal $\gamma$ and hence determine the best topology for analyzing a given dataset. The procedures proposed in this work open the door for more in-depth study of the $L^\gamma$-graphs and what determines their optimal topology. They also pave the way towards the extension of other standard clustering tools, such as Unsupervised Learning via Spectral Clustering, to exploit these richer topologies.


\appendix
\section{Proofs}

\subsection{Proof of Lemma \ref{GCR-Interpretation.lemma}}\label{GCR-Interpretation.proof}
\begin{proof}
Let $r = (\mathcal{A}_{out}(S) - \mathcal{D}_{out}(S)) / (\mathcal{A}_{in}(S) - \mathcal{D}_{in}(S))$. It is easy to show that $h_S^{(\gamma)} = r/(r+1)$, which is monotonocally increasing with $r$. Thus, the task is the partition that minimizes $r$ and consequently $h^{(\gamma)}_S$.
\end{proof}

\subsection{Proof of Lemma \ref{LgPR-solution.lemma}}\label{LgPR-solution.proof}
\begin{proof} It suffices to show the positive semi-definiteness of the functional and to apply the first order optimality condition. Let $\tilde{f} = Q^T D_{\gamma}^{-1} f$. Then, the left term satisfies $\sum_{j} \lambda_{j}^{\gamma} \tilde{f}_j^2 \geq 0$. It can be shown that $[D_{\gamma}]_{uu} = \sum_{j} Q_{uj}^2 \lambda_j^{\gamma} \geq 0$ granting the right term satisfies $\sum_u (f_u - y_u)^2 / [D_{\gamma}]_{uu} \geq 0$. Now, computing the derivative of the functional with respect to $f$ and equaling to $0$ leads to: $L^{\gamma} D_{\gamma}^{-1}f + \mu (f - y) = 0$. The lemma is proved after isolating $f$.
\end{proof}

\subsection{Proof of Lemma \ref{LgPR-properties.lemma}}\label{LgPR-properties.proof}
\begin{proof}
	From the demonstration of Lemma (\ref{LgPR-solution.lemma}) we have that $L^{\gamma}D_{\gamma}^{-1} f + \mu (f - y) = 0$. Then, $\mathbbm{1}^T L^{\gamma}D_{\gamma}^{-1} + \mu \mathbbm{1}^T f = \mu \mathbbm{1}^T y$. Since $\mathbbm{1}^T L^{\gamma} = 0$ we have that $\mathbbm{1}^T f = \mathbbm{1}^T y$, proving (\textit{i}). We prove property (\textit{iii}) using the same expression. We only develop the case $\mu \to 0$ since the case $\mu \to \infty$ follows the same steps: taking $\lim_{\mu \to 0} \left\{ L^{\gamma}D_{\gamma}^{-1}f + \mu (f-y) = 0 \right\}$ leads to $L^{\gamma}D_{\gamma}^{-1} f = 0$, whose solution is proportional to $\pi_{\gamma} = D_{\gamma}\mathbbm{1}/vol_{\gamma}(G)$. Lastly, we prove (\textit{ii}) by noting that the operator $L^{\gamma}D_{\gamma}^{-1}$ has a positive real spectrum as it is similar to $D_{\gamma}^{-1/2} L^{\gamma} D_{\gamma}^{-1/2}$ which is positive semi-definite. Thus, we can use the inverse Laplace transform of the resolvent $(L^{\gamma}D_{\gamma}^{-1} + \mu \mathbb{I})^{-1} = \int_{0}^{\infty} e^{-t} e^{ -tL^{\gamma}D_{\gamma}^{-1}/\mu} dt$, which, after using its Taylor expansion, allows to rewrite the PageRank solution as $f = \sum_{k = 0}^{\infty} \frac{(-1)^k}{\mu^k} \left( L^{\gamma}D_{\gamma}^{-1} \right)^k y$.
	If $y = \pi_{\gamma}$, the previous equation is only non-zero for $k = 0$, proving (\textit{ii}).  
\end{proof}

\subsection{Proof of Lemma \ref{LgPR-bound.lemma}}\label{LgPR-bound.proof}
\begin{proof}
	Let $y = D_{\gamma}\mathbbm{1}_S/vol_{\gamma}(S)$. Using (\ref{LgPR-solution.eq}) we can see that 
	\begin{equation} 
		\mathbbm{1}_{S^c}^T f = \sum_{u \in S} \frac{\left[ D_{\gamma} \right]_{uu}}{vol_{\gamma}(S)} \mathbbm{1}_{S^c}^T \left[ \mu \left( L^{\gamma}D_{\gamma}^{-1} + \mu \mathbb{I} \right)^{-1} \delta_u \right],
	\end{equation}
	showing that $\mathbbm{1}_{S^c}^T f$ can be interpreted as $\mathbb{E}\left[ f(S^c) \right]$ when labels are selected with probability proportional to their generalized degree in $S$. Using the fact that 
	\begin{equation} 
		\left( L^{\gamma}D_{\gamma}^{-1} + \mu \mathbb{I} \right)^{-1} \left( L^{\gamma}D_{\gamma}^{-1} + \mu \mathbb{I} \right) = \mathbb{I},
	\end{equation}
	we express 
	\begin{equation} 
		f = \left( \mathbb{I} - \frac{1}{\mu} L^{\gamma}D_{\gamma}^{-1} + \frac{1}{\mu} L^{\gamma}D_{\gamma}^{-1} \left( L^{\gamma} D_{\gamma}^{-1} + \mu \mathbb{I} \right)^{-1} L^{\gamma}D_{\gamma}^{-1} \right)y.
	\end{equation}
	The upper bound is thus obtained by substituting $y$ and summing over $S$.
	\begin{align} 
		\nonumber \mathbbm{1}_S^T f &= \frac{ \mathbbm{1}_S^T D_{\gamma} \mathbbm{1}_S }{vol_{\gamma}(S)}  - \frac{\mathbbm{1}_S^T L^{\gamma} \mathbbm{1}_{S}}{\mu~vol_{\gamma}(S)}  + \frac{ \mathbbm{1}_S^T L^{\gamma} \left( L^{\gamma}  + \mu D_{\gamma} \right)^{-1} L^{\gamma} \mathbbm{1}_S }{\mu~vol_{\gamma}(S)}, \\
		\nonumber & \geq \frac{ \mathbbm{1}_S^T D_{\gamma} \mathbbm{1}_S }{vol_{\gamma}(S)}  - \frac{\mathbbm{1}_S^T L^{\gamma} \mathbbm{1}_{S}}{\mu~vol_{\gamma}(S)}, \\
		& = 1 - \frac{h_{S}^{(\gamma)}}{\mu} .
	\end{align}
	Employing property (\textit{i}) from Lemma \ref{LgPR-properties.lemma} finishes the proof.
\end{proof}

\subsection{Proof of Lemma \ref{LgPR-sharpdrop-ineq.lemma}}\label{LgPR-sharpdrop-ineq.proof}
\begin{proof}
We only show the proof of the lower bound as the upper bound follows a similar derivation. We recast (\ref{LgPR-solution.lemma}) as $L^{\gamma}D_{\gamma}^{-1}f = \mu\left( y - f \right)$. Thus, the set $S_j$ satisfies: 
\begin{flalign}
    \mu \left((y(S_j) - f(S_j) \right) &= \mathbbm{1}_{S_j}^T L^{\gamma} D_\gamma^{-1} f \notag\\
    &= \mathbbm{1}_{S_j}^T L^{\gamma} q \notag\\
    &= \sum_{u \in S_j, v \in S_j^c} \left[ W_{\gamma} \right]_{uv} (q_u - q_v) \notag\\
    &= \sum_{u \in S_j, v \in S_j^c} |\left[ W_{\gamma}^+ \right]_{uv}|(q_u - q_v) - \sum_{u \in S_j, v \in S_j^c} |\left[ W_{\gamma}^- \right]_{uv}|(q_u - q_v) \notag\\ 
    &\qquad+ \sum_{u \in S_j, v \in S_j^c} |\left[ W_{\gamma}^+ \right]_{uv}| (q_{j} - q_{j+1})  - \sum_{u \in S_j, v \in S_j^c} |\left[ W_{\gamma}^+ \right]_{uv}|(q_j - q_{j+1}) \notag\\
    &\qquad+ \sum_{u \in S_j, v \in S_j^c} |\left[ W_{\gamma}^- \right]_{uv}|(q_j - q_{j+1}) - \sum_{u \in S_j, v \in S_j^c} |\left[ W_{\gamma}^- \right]_{uv}|(q_{j} - q_{j+1}) \notag\\
    &\geq (q_j - q_{j+1}) \left( 2\mathcal{A}_{out}(S_j) + \mathcal{D}_{out}(S_j) \right) \notag\\ 
    &\qquad - (q_1 - q_N) \left( 2\mathcal{D}_{out}(S_j) + \mathcal{A}_{out}(S_j) \right)&&
\end{flalign}
Re-ordering terms finishes the proof.
\end{proof}

\subsection{Proof of Theorem \ref{SBM-L2.theorem}}\label{SBM-L2.theorem.proof}
\begin{proof}
	Let $n = |S|$. For $u, v \in S$ and $w \in S^c$ the Planted Partition satisfies $\sum_{v} W_{uv} \sim B(n-1, p_{in})$  and $\sum_{w} W_{uw} \sim B(n, p_{out})$. The key step in the proof is to show that, in the limit $n \to \infty$, $\mathbb{E}\left[ h_S^{(1)} \right] =  \mathbb{E}\left[ \frac{\mathbbm{1}_S^T L \mathbbm{1}_S}{vol(S)}\right] = \frac{ \mathbb{E}\left[ \mathbbm{1}_S^T L \mathbbm{1}_S \right]}{\mathbb{E} \left[ vol(S) \right]} $, and the same for $h_S^{(2)}$. By application of the Chebyshev inequality we have that 
	\begin{equation}
		Pr \left( d_u - \mathbb{E}\left[ d_u \right] \geq \mathbb{E}\left[ d_u \right] \right) \leq \frac{var(d_u)}{var(d_u) + \mathbb{E}\left[d_u \right]^2}= \mathcal{O}(n^{-1}).
	\end{equation}
	Thus, in the limit of $n \to \infty$ we can establish the inequality $d_u < 2\mathbb{E}\left[ d_u \right]$ and further that $vol(S) < 2 \mathbb{E}\left[vol(S)\right]$. This latter allows to express $\mathbb{E}\left[ h_S^{(1)} \right]$ as follows \cite{Rice.2015.expected}:
	\begin{align}
 	\nonumber	\mathbb{E}\left[ h_S^{(1)} \right] &= \mathbb{E}\left[ \frac{\mathbbm{1}_S^T L \mathbbm{1}_S}{vol(S)}\right] \\
	\nonumber 	&=  \frac{ \mathbb{E}\left[ \mathbbm{1}_S^T L \mathbbm{1}_S \right]}{\mathbb{E} \left[ vol(S) \right]} + \sum_{i = 1}^{\infty} (-1)^i \frac{\mathbb{E}[ \mathbbm{1}_S^T L \mathbbm{1}_S] \llangle ^i vol(S) \rrangle + \llangle \mathbbm{1}_S^T L \mathbbm{1}_S, ^i vol(S) \rrangle}{\mathbb{E}\left[ vol(S) \right]^{i+1}} \\
	\nonumber	&= \frac{ \mathbb{E}\left[ \mathbbm{1}_S^T L \mathbbm{1}_S \right]}{\mathbb{E} \left[ vol(S) \right]} +\sum_{i = 1}^{\infty} (-1)^i \frac{\mathbb{E}\left[ \mathbbm{1}_S^T L \mathbbm{1}_S (vol(S) - \mathbb{E}[vol(S)])^i\right] }{\mathbb{E}[vol(S)]^{i+1}} \\
	\nonumber	&= \frac{ \mathbb{E}\left[ \mathbbm{1}_S^T L \mathbbm{1}_S \right]}{\mathbb{E} \left[ vol(S) \right]} + \sum_{i = 1}^{\infty} (-1)^i \mathbb{E}\left[\frac{\mathbbm{1}_S^T L \mathbbm{1}_S}{\mathbb{E}[vol(S)]}  \left( \frac{ vol(S)}{\mathbb{E}[vol(S)]} - 1\right)^i \right] \\
		&= \frac{ \mathbb{E}\left[ \mathbbm{1}_S^T L \mathbbm{1}_S \right]}{\mathbb{E} \left[ vol(S) \right]} + \sum_{i = 1}^{\infty} (-1)^i c_i 
	\end{align}
	where $\llangle a,^i b \rrangle = \mathbb{E}\left[ (a - \mathbb{E}[a]) (b - \mathbb{E}[b])^i \right]$. The fact that $vol(S) < 2 \mathbb{E}\left[vol(S)\right]$ and the monotonicity of the expected value imply that the sequence $\sum_{i} |c_i|$ decreases monotonically. Also, it can be shown that its dominant term: $c_1 = \mathcal{O}(n^{-2})$. Replacing the expectations and evaluating the limit leads to
	\begin{equation}
		\lim_{n \to \infty} \mathbb{E} \left[ h_S^{(1)} \right] = \frac{p_{out}}{p_{in} + p_{out}}.
	\end{equation}	
	The case of $\mathbb{E}[ h_S^{(2)} ]$ follows a similar derivation. Since $\left[ D_{2} \right]_{uu} = d_u^2 + d_u$, the Jensen inequality implies that $\left[ D_2\right]_{uu} < 2 \mathbb{E}\left[ [D_{2}]_{uu} \right]$ and consequently that $vol_2(S) < 2 \mathbb{E}\left[ vol_2(S) \right]$. Thus, we cast
	\begin{align}
	\nonumber	\mathbb{E}\left[ h_S^{(2)} \right] &= \mathbb{E}\left[ \frac{\mathbbm{1}_S^T L^2 \mathbbm{1}_S}{vol_2(S)}\right]  \\
	\nonumber	&= \frac{ \mathbb{E}\left[ \mathbbm{1}_S^T L^2 \mathbbm{1}_S \right]}{\mathbb{E} \left[ vol_2(S) \right]} + \sum_{i = 1}^{\infty} (-1)^i \mathbb{E}\left[\frac{\mathbbm{1}_S^T L^2 \mathbbm{1}_S}{\mathbb{E}[vol_2(S)]}  \left( \frac{ vol_2(S)}{\mathbb{E}[vol_2(S)]} - 1\right)^i \right] \\
		&= \frac{ \mathbb{E}\left[ \mathbbm{1}_S^T L^2 \mathbbm{1}_S \right]}{\mathbb{E} \left[ vol_2(S) \right]} + \sum_{i = 1}^{\infty} (-1)^i c^{(2)}_i 
	\end{align}	
	Let the random variable $O_u = \sum_{w \in S^c} W_{uw}$. Then we have that $\mathbbm{1}_S^T L^2 \mathbbm{1}_S = 2 \sum_{u \in S} \left( O_u \right)^2$. This fact, in addition to $vol_{2}(S) = \sum_{u\in S}d_u^2 + d_u$, allow to show that the sequence $\sum_{i} |c^{(2)}_i|$ is monotonically decreasing with $c^{(2)}_1 = \mathcal{O}(n^{-1})$. Replacing the expectations and evaluating the limit leads to 
	\begin{equation}
		\lim_{n \to \infty} \mathbb{E} \left[ h_S^{(2)} \right] = 2 \left( \frac{p_{out}}{p_{in}+p_{out}} \right)^2
	\end{equation}
\end{proof}

\subsection{Proof of Corollary \ref{SBM-L2.corollary}}\label{SBM-L2.corollary.proof}
\begin{proof}
	Let $p_{in} = p_{out} + \epsilon$ and assume that $h_{S}^{(1)} \geq h_S^{(2)}$. Thus $p_{out}/(p_{in} + p_{out}) \geq 2 (p_{out}/(p_{in} + p_{out}))^2$, which can be further simplified to $1 \geq 2 p_{out}/( 2 p_{out} + \epsilon)$. We observe that such expression holds for $\epsilon \geq 0$ and equality occurs when $\epsilon = 0$.
\end{proof}

\subsection{Proof of Proposition \ref{Condition-L2.proposition}}\label{Condition-L2.proof}
\begin{proof}
	We search a condition on $S$ that permits $\frac{\mathbbm{1}_S^T L \mathbbm{1}_S}{\mathbbm{1}_S^T D \mathbbm{1}_S} \geq \frac{\mathbbm{1}_S^T L^2 \mathbbm{1}_S}{\mathbbm{1}_S^T D_2 \mathbbm{1}_S} $, or equivalently, that satisfies the inequality $\frac{\mathbbm{1}_S^T D_2 \mathbbm{1}_S}{\mathbbm{1}_S^T D \mathbbm{1}_S}-\frac{\mathbbm{1}_S^T L^2 \mathbbm{1}_S}{\mathbbm{1}_S^T L \mathbbm{1}_S} \geq 0$. We have
\begin{align}
	\nonumber \frac{\mathbbm{1}_S^T D_2 \mathbbm{1}_S}{\mathbbm{1}_S^T D \mathbbm{1}_S}-\frac{\mathbbm{1}_S^T L^2 \mathbbm{1}_S}{\mathbbm{1}_S^T L \mathbbm{1}_S} &\geq \frac{\mathbbm{1}_S^T D^2 \mathbbm{1}_S}{\mathbbm{1}_S^T D \mathbbm{1}_S}-\frac{\mathbbm{1}_S^T L^2 \mathbbm{1}_S}{\mathbbm{1}_S^T L \mathbbm{1}_S} \\
	\nonumber &\geq  \frac{\mathbbm{1}_S^T D^2 \mathbbm{1}_S}{\mathbbm{1}_S^T D \mathbbm{1}_S} - \left( \max_{u \in S} \sum_{w \in S^c} W_{uw} + \max_{\ell \in S^c} \sum_{v \in S} W_{\ell v} \right) \\
	\nonumber &\geq \frac{\mathbbm{1}_S^T D \mathbbm{1}_S}{\mathbbm{1}_S^T \mathbbm{1}_S} - \left( \max_{u \in S} \sum_{w \in S^c} W_{uw} + \max_{\ell \in S^c} \sum_{v \in S} W_{\ell v} \right)\\
    &= \frac{vol(S)}{|S|} - \left( \max_{u \in S} \sum_{w \in S^c} W_{uw} + \max_{\ell \in S^c} \sum_{v \in S} W_{\ell v} \right),
\end{align}
where we have used Lehmer’s and Holder’s inequalities and that $\mathbbm{1}_S^T L^2\mathbbm{1}_S = \sum_{u \in S}\left( \sum_{w \in S^c} W_{uw} \right)^2 + \sum_{\ell \in S^c} \left( \sum_{v \in S} W_{\ell v} \right)^2 $. Thus, it is sufficient that $S$ satisfies
\begin{equation}
    \frac{vol(S)}{|S|} - \left( \max_{u \in S} \sum_{w \in S^c} W_{uw} + \max_{\ell \in S^c} \sum_{v \in S} W_{\ell v} \right) \geq 0
\end{equation}
\end{proof}


\begin{backmatter}
\section*{Availability of data and material}
All data generated or analysed during this study are included the following articles \cite{Lecun.1998.gradient, Hond.1997.distinctive, Greene.2006.practical,phoneme}. The code to replicate the results is available in the GitHub repository, \href{https://github.com/estbautista/Lgamma-PageRank_Paper}{https://github.com/estbautista/Lgamma-PageRank{\_}Paper}.

\section*{Competing interests}
  The authors declare that they have no competing interests.
  
\section*{Funding}
This work was supported by CONACyT and the Labex MILyon
  
\section*{Author's contributions}
EB, PA and PG participated equally in designing and developing the project, and in writing the paper. 

\section*{Acknowledgements}
Not applicable


\bibliographystyle{bmc-mathphys} 
\bibliography{bmc_article}      


\begin{thebibliography}{30}
\ifx \bisbn   \undefined \def \bisbn  #1{ISBN #1}\fi
\ifx \binits  \undefined \def \binits#1{#1}\fi
\ifx \bauthor  \undefined \def \bauthor#1{#1}\fi
\ifx \batitle  \undefined \def \batitle#1{#1}\fi
\ifx \bjtitle  \undefined \def \bjtitle#1{#1}\fi
\ifx \bvolume  \undefined \def \bvolume#1{\textbf{#1}}\fi
\ifx \byear  \undefined \def \byear#1{#1}\fi
\ifx \bissue  \undefined \def \bissue#1{#1}\fi
\ifx \bfpage  \undefined \def \bfpage#1{#1}\fi
\ifx \blpage  \undefined \def \blpage #1{#1}\fi
\ifx \burl  \undefined \def \burl#1{\textsf{#1}}\fi
\ifx \doiurl  \undefined \def \doiurl#1{\textsf{#1}}\fi
\ifx \betal  \undefined \def \betal{\textit{et al.}}\fi
\ifx \binstitute  \undefined \def \binstitute#1{#1}\fi
\ifx \binstitutionaled  \undefined \def \binstitutionaled#1{#1}\fi
\ifx \bctitle  \undefined \def \bctitle#1{#1}\fi
\ifx \beditor  \undefined \def \beditor#1{#1}\fi
\ifx \bpublisher  \undefined \def \bpublisher#1{#1}\fi
\ifx \bbtitle  \undefined \def \bbtitle#1{#1}\fi
\ifx \bedition  \undefined \def \bedition#1{#1}\fi
\ifx \bseriesno  \undefined \def \bseriesno#1{#1}\fi
\ifx \blocation  \undefined \def \blocation#1{#1}\fi
\ifx \bsertitle  \undefined \def \bsertitle#1{#1}\fi
\ifx \bsnm \undefined \def \bsnm#1{#1}\fi
\ifx \bsuffix \undefined \def \bsuffix#1{#1}\fi
\ifx \bparticle \undefined \def \bparticle#1{#1}\fi
\ifx \barticle \undefined \def \barticle#1{#1}\fi
\ifx \bconfdate \undefined \def \bconfdate #1{#1}\fi
\ifx \botherref \undefined \def \botherref #1{#1}\fi
\ifx \url \undefined \def \url#1{\textsf{#1}}\fi
\ifx \bchapter \undefined \def \bchapter#1{#1}\fi
\ifx \bbook \undefined \def \bbook#1{#1}\fi
\ifx \bcomment \undefined \def \bcomment#1{#1}\fi
\ifx \oauthor \undefined \def \oauthor#1{#1}\fi
\ifx \citeauthoryear \undefined \def \citeauthoryear#1{#1}\fi
\ifx \endbibitem  \undefined \def \endbibitem {}\fi
\ifx \bconflocation  \undefined \def \bconflocation#1{#1}\fi
\ifx \arxivurl  \undefined \def \arxivurl#1{\textsf{#1}}\fi
\csname PreBibitemsHook\endcsname

\bibitem{Avrachenkov.2012.classification}
\begin{bchapter}
\bauthor{\bsnm{Avrachenkov}, \binits{K.}},
\bauthor{\bsnm{Gon{\c c}alves}, \binits{P.}},
\bauthor{\bsnm{Legout}, \binits{A.}},
\bauthor{\bsnm{Sokol}, \binits{M.}}:
\bctitle{Classification of Content and Users in BitTorrent by Semi-supervised
  Learning Methods}.
In: \bbtitle{International Wireless Communications and Mobile Computing
  Conference (3rd International Workshop on Traffic Analysis and
  Classification)},
\blocation{Cyprus}
(\byear{2012}).
\bcomment{Best paper award}
\end{bchapter}
\endbibitem

\bibitem{Subramanya.2008.soft}
\begin{bchapter}
\bauthor{\bsnm{Subramanya}, \binits{A.}},
\bauthor{\bsnm{Bilmes}, \binits{J.}}:
\bctitle{Soft-supervised learning for text classification}.
In: \bbtitle{Proceedings of the Conference on Empirical Methods in Natural
  Language Processing}.
\bsertitle{EMNLP '08},
pp. \bfpage{1090}--\blpage{1099}.
\bpublisher{Association for Computational Linguistics},
\blocation{Stroudsburg, PA, USA}
(\byear{2008})
\end{bchapter}
\endbibitem

\bibitem{Zhao.2014.compact}
\begin{barticle}
\bauthor{\bsnm{{Zhao}}, \binits{M.}},
\bauthor{\bsnm{{Chan}}, \binits{R.H.M.}},
\bauthor{\bsnm{{Chow}}, \binits{T.W.S.}},
\bauthor{\bsnm{{Tang}}, \binits{P.}}:
\batitle{Compact graph based semi-supervised learning for medical diagnosis in
  alzheimer’s disease}.
\bjtitle{IEEE Signal Processing Letters}
\bvolume{21}(\bissue{10}),
\bfpage{1192}--\blpage{1196}
(\byear{2014}).
doi:\doiurl{10.1109/LSP.2014.2329056}
\end{barticle}
\endbibitem

\bibitem{Fontugne.2019.BGP}
\begin{bchapter}
\bauthor{\bsnm{Fontugne}, \binits{R.}},
\bauthor{\bsnm{Bautista}, \binits{E.}},
\bauthor{\bsnm{Petrie}, \binits{C.}},
\bauthor{\bsnm{Nomura}, \binits{Y.}},
\bauthor{\bsnm{Abry}, \binits{P.}},
\bauthor{\bsnm{Gon{\c c}alves}, \binits{P.}},
\bauthor{\bsnm{Fukuda}, \binits{K.}},
\bauthor{\bsnm{Aben}, \binits{E.}}:
\bctitle{{BGP Zombies: an Analysis of Beacons Stuck Routes}}.
In: \bbtitle{{PAM 2019 - 20th Passive and Active Measurements Conference}},
\bconflocation{Puerto Varas, Chile},
pp. \bfpage{1}--\blpage{13}
(\byear{2019}).
\burl{https://hal.inria.fr/hal-01970596}
\end{bchapter}
\endbibitem

\bibitem{Chung.2010.pagerank}
\begin{botherref}
\oauthor{\bsnm{Chung}, \binits{F.}}:
Pagerank as a discrete green's function.
Geometry and Analysis I ALM
(2010)
\end{botherref}
\endbibitem

\bibitem{Avrachenkov.2018.mean}
\begin{barticle}
\bauthor{\bsnm{Avrachenkov}, \binits{K.}},
\bauthor{\bsnm{Kadavankandy}, \binits{A.}},
\bauthor{\bsnm{Litvak}, \binits{N.}}:
\batitle{{Mean Field Analysis of Personalized PageRank with Implications for
  Local Graph Clustering}}.
\bjtitle{{Journal of Statistical Physics}}
\bvolume{173}(\bissue{3-4}),
\bfpage{895}--\blpage{916}
(\byear{2018}).
doi:\doiurl{10.1007/s10955-018-2099-5}
\end{barticle}
\endbibitem

\bibitem{Litvak.2009.characterization}
\begin{bchapter}
\bauthor{\bsnm{Litvak}, \binits{N.}},
\bauthor{\bsnm{Scheinhardt}, \binits{W.}},
\bauthor{\bsnm{Volkovich}, \binits{Y.}},
\bauthor{\bsnm{Zwart}, \binits{B.}}:
\bctitle{Characterization of tail dependence for in-degree and pagerank}.
In: \beditor{\bsnm{Avrachenkov}, \binits{K.}},
\beditor{\bsnm{Donato}, \binits{D.}},
\beditor{\bsnm{Litvak}, \binits{N.}} (eds.)
\bbtitle{Algorithms and Models for the Web-Graph},
pp. \bfpage{90}--\blpage{103}.
\bpublisher{Springer},
\blocation{Berlin, Heidelberg}
(\byear{2009})
\end{bchapter}
\endbibitem

\bibitem{Chung.2007.four}
\begin{bchapter}
\bauthor{\bsnm{Chung}, \binits{F.}}:
\bctitle{Four cheeger-type inequalities for graph partitioning algorithms}.
In: \bbtitle{Proceedings of ICCM}
(\byear{2007})
\end{bchapter}
\endbibitem

\bibitem{Avrachenkov.2008.pagerank}
\begin{bchapter}
\bauthor{\bsnm{Avrachenkov}, \binits{K.}},
\bauthor{\bsnm{Dobrynin}, \binits{V.}},
\bauthor{\bsnm{Nemirovsky}, \binits{D.}},
\bauthor{\bsnm{Pham}, \binits{S.K.}},
\bauthor{\bsnm{Smirnova}, \binits{E.}}:
\bctitle{Pagerank based clustering of hypertext document collections}.
In: \bbtitle{Proceedings of the 31st Annual International ACM SIGIR Conference
  on Research and Development in Information Retrieval},
pp. \bfpage{873}--\blpage{874}
(\byear{2008})
\end{bchapter}
\endbibitem

\bibitem{Graham.2009.distributing}
\begin{barticle}
\bauthor{\bsnm{Graham}, \binits{F.C.}},
\bauthor{\bsnm{Horn}, \binits{P.}},
\bauthor{\bsnm{Tsiatas}, \binits{A.}}:
\batitle{Distributing antidote using pagerank vectors}.
\bjtitle{Internet Mathematics}
\bvolume{6},
\bfpage{237}--\blpage{254}
(\byear{2009})
\end{barticle}
\endbibitem

\bibitem{Andersen.2007.using}
\begin{barticle}
\bauthor{\bsnm{Andersen}, \binits{R.}},
\bauthor{\bsnm{R.~K.~Chung}, \binits{F.}},
\bauthor{\bsnm{J.~Lang}, \binits{K.}}:
\batitle{Using pagerank to locally partition a graph}.
\bjtitle{Internet Mathematics}
\bvolume{4},
\bfpage{35}--\blpage{64}
(\byear{2007}).
doi:\doiurl{10.1080/15427951.2007.10129139}
\end{barticle}
\endbibitem

\bibitem{Andersen.2007.detecting}
\begin{bchapter}
\bauthor{\bsnm{Andersen}, \binits{R.}},
\bauthor{\bsnm{Chung}, \binits{F.}}:
\bctitle{Detecting sharp drops in pagerank and a simplified local partitioning
  algorithm}.
In: \beditor{\bsnm{Cai}, \binits{J.-Y.}},
\beditor{\bsnm{Cooper}, \binits{S.B.}},
\beditor{\bsnm{Zhu}, \binits{H.}} (eds.)
\bbtitle{Theory and Applications of Models of Computation},
pp. \bfpage{1}--\blpage{12}.
\bpublisher{Springer},
\blocation{Berlin, Heidelberg}
(\byear{2007})
\end{bchapter}
\endbibitem

\bibitem{Zhou.2003.learning}
\begin{bchapter}
\bauthor{\bsnm{Zhou}, \binits{D.}},
\bauthor{\bsnm{Bousquet}, \binits{O.}},
\bauthor{\bsnm{Lal}, \binits{T.N.}},
\bauthor{\bsnm{Weston}, \binits{J.}},
\bauthor{\bsnm{Sch\"{o}lkopf}, \binits{B.}}:
\bctitle{Learning with local and global consistency}.
In: \beditor{\bsnm{Thrun}, \binits{S.}},
\beditor{\bsnm{Saul}, \binits{L.K.}},
\beditor{\bsnm{Sch\"{o}lkopf}, \binits{B.}} (eds.)
\bbtitle{Advances in Neural Information Processing Systems 16},
pp. \bfpage{321}--\blpage{328}
(\byear{2004}).
\burl{http://papers.nips.cc/paper/2506-learning-with-local-and-global-consistency.pdf}
\end{bchapter}
\endbibitem

\bibitem{Zhou.2007.spectral}
\begin{bchapter}
\bauthor{\bsnm{Zhou}, \binits{D.}},
\bauthor{\bsnm{Burges}, \binits{C.J.C.}}:
\bctitle{Spectral clustering and transductive learning with multiple views}.
In: \bbtitle{Proceedings of the 24th International Conference on Machine
  Learning}.
\bsertitle{ICML '07},
pp. \bfpage{1159}--\blpage{1166}.
\bpublisher{ACM},
\blocation{New York, NY, USA}
(\byear{2007}).
doi:\doiurl{10.1145/1273496.1273642}
\end{bchapter}
\endbibitem

\bibitem{Avrachenkov.2012.generalized}
\begin{bchapter}
\bauthor{\bsnm{Avrachenkov}, \binits{K.}},
\bauthor{\bsnm{Gon{\c c}alves}, \binits{P.}},
\bauthor{\bsnm{Mishenin}, \binits{A.}},
\bauthor{\bsnm{Sokol}, \binits{M.}}:
\bctitle{Generalized Optimization Framework for Graph-based Semi-supervised
  Learning}.
In: \bbtitle{SIAM Data Mining}
(\byear{2012})
\end{bchapter}
\endbibitem

\bibitem{Zhou.2011.semi}
\begin{bchapter}
\bauthor{\bsnm{Zhou}, \binits{X.}},
\bauthor{\bsnm{Belkin}, \binits{M.}}:
\bctitle{Semi-supervised learning by higher order regularization}.
In: \beditor{\bsnm{Gordon}, \binits{G.}},
\beditor{\bsnm{Dunson}, \binits{D.}},
\beditor{\bsnm{Dudík}, \binits{M.}} (eds.)
\bbtitle{Proceedings of the Fourteenth International Conference on Artificial
  Intelligence and Statistics}.
\bsertitle{Proceedings of Machine Learning Research},
vol. \bseriesno{15},
pp. \bfpage{892}--\blpage{900}.
\bpublisher{PMLR},
\blocation{Fort Lauderdale, FL, USA}
(\byear{2011}).
\burl{http://proceedings.mlr.press/v15/zhou11b/zhou11b.pdf}
\end{bchapter}
\endbibitem

\bibitem{Tsiatas.2012.diffusion}
\begin{botherref}
\oauthor{\bsnm{Tsiatas}, \binits{A.}}:
Diffusion and clustering on large graphs.
PhD thesis,
University of California at San Diego,
La Jolla, CA, USA
(2012)
\end{botherref}
\endbibitem

\bibitem{Sokol.2014.graph}
\begin{botherref}
\oauthor{\bsnm{Sokol}, \binits{M.}}:
Graph-based semi-supervised learning methods and quick detection of central
  nodes.
PhD thesis,
Universit{\'e} de Nice, Ecole Doctorale STIC,
Inria Sophia Antipolis, Maestro
(April 2014)
\end{botherref}
\endbibitem

\bibitem{Riascos.2014.fractional}
\begin{barticle}
\bauthor{\bsnm{P\'erez~Riascos}, \binits{A.}},
\bauthor{\bsnm{Mateos}, \binits{J.}}:
\batitle{Fractional dynamics on networks: Emergence of anomalous diffusion and
  l\'evy flights}.
\bjtitle{Physical Review E}
\bvolume{90},
\bfpage{032809}
(\byear{2014}).
doi:\doiurl{10.1103/PhysRevE.90.032809}
\end{barticle}
\endbibitem

\bibitem{DeNigris.2017.fractional}
\begin{bchapter}
\bauthor{\bsnm{{de Nigris}}, \binits{S.}},
\bauthor{\bsnm{{Bautista}}, \binits{E.}},
\bauthor{\bsnm{{Abry}}, \binits{P.}},
\bauthor{\bsnm{{Avrachenkov}}, \binits{K.}},
\bauthor{\bsnm{{Goncalves}}, \binits{P.}}:
\bctitle{Fractional graph-based semi-supervised learning}.
In: \bbtitle{2017 25th European Signal Processing Conference (EUSIPCO)},
pp. \bfpage{356}--\blpage{360}
(\byear{2017}).
doi:\doiurl{10.23919/EUSIPCO.2017.8081228}
\end{bchapter}
\endbibitem

\bibitem{Bautista.2017.Levy}
\begin{bchapter}
\bauthor{\bsnm{Bautista}, \binits{E.}},
\bauthor{\bsnm{De~Nigris}, \binits{S.}},
\bauthor{\bsnm{Abry}, \binits{P.}},
\bauthor{\bsnm{Avrachenkov}, \binits{K.}},
\bauthor{\bsnm{Gon{\c c}alves}, \binits{P.}}:
\bctitle{{L{\'e}vy Flights for Graph Based Semi-Supervised Classification. }}.
In: \bbtitle{{26th Colloquium GRETSI}}.
\bsertitle{GRETSI, 2017 - Proceeding of the 26th colloquium},
\bconflocation{Juan-Les-Pins, France}
(\byear{2017})
\end{bchapter}
\endbibitem

\bibitem{Shuman.2018.distributed}
\begin{barticle}
\bauthor{\bsnm{{Shuman}}, \binits{D.I.}},
\bauthor{\bsnm{{Vandergheynst}}, \binits{P.}},
\bauthor{\bsnm{{Kressner}}, \binits{D.}},
\bauthor{\bsnm{{Frossard}}, \binits{P.}}:
\batitle{Distributed signal processing via chebyshev polynomial approximation}.
\bjtitle{IEEE Transactions on Signal and Information Processing over Networks}
\bvolume{4}(\bissue{4}),
\bfpage{736}--\blpage{751}
(\byear{2018}).
doi:\doiurl{10.1109/TSIPN.2018.2824239}
\end{barticle}
\endbibitem

\bibitem{Mossel.2015.reconstruction}
\begin{barticle}
\bauthor{\bsnm{Mossel}, \binits{E.}},
\bauthor{\bsnm{Neeman}, \binits{J.}},
\bauthor{\bsnm{Sly}, \binits{A.}}:
\batitle{Reconstruction and estimation in the planted partition model}.
\bjtitle{Probability Theory and Related Fields}
\bvolume{162}(\bissue{3}),
\bfpage{431}--\blpage{461}
(\byear{2015}).
doi:\doiurl{10.1007/s00440-014-0576-6}
\end{barticle}
\endbibitem

\bibitem{Zhang.2014.phase}
\begin{botherref}
\oauthor{\bsnm{Zhang}, \binits{P.}},
\oauthor{\bsnm{Moore}, \binits{C.}},
\oauthor{\bsnm{Zdeborova}, \binits{L.}}:
Phase transitions in semisupervised clustering of sparse networks.
Physical review. E, Statistical, nonlinear, and soft matter physics
\textbf{90}
(2014).
doi:\doiurl{10.1103/PhysRevE.90.052802}
\end{botherref}
\endbibitem

\bibitem{Matthews.1975.comparison}
\begin{barticle}
\bauthor{\bsnm{Matthews}, \binits{B.W.}}:
\batitle{Comparison of the predicted and observed secondary structure of t4
  phage lysozyme}.
\bjtitle{Biochimica et Biophysica Acta (BBA) - Protein Structure}
\bvolume{405}(\bissue{2}),
\bfpage{442}--\blpage{451}
(\byear{1975}).
doi:\doiurl{10.1016/0005-2795(75)90109-9}
\end{barticle}
\endbibitem

\bibitem{Lecun.1998.gradient}
\begin{barticle}
\bauthor{\bsnm{{Lecun}}, \binits{Y.}},
\bauthor{\bsnm{{Bottou}}, \binits{L.}},
\bauthor{\bsnm{{Bengio}}, \binits{Y.}},
\bauthor{\bsnm{{Haffner}}, \binits{P.}}:
\batitle{Gradient-based learning applied to document recognition}.
\bjtitle{Proceedings of the IEEE}
\bvolume{86}(\bissue{11}),
\bfpage{2278}--\blpage{2324}
(\byear{1998}).
doi:\doiurl{10.1109/5.726791}
\end{barticle}
\endbibitem

\bibitem{Hond.1997.distinctive}
\begin{bchapter}
\bauthor{\bsnm{Hond}, \binits{D.}},
\bauthor{\bsnm{Spacek}, \binits{L.}}:
\bctitle{Distinctive descriptions for face processing.}
In: \beditor{\bsnm{Clark}, \binits{A.F.}} (ed.)
\bbtitle{BMVC}
(\byear{1997})
\end{bchapter}
\endbibitem

\bibitem{Greene.2006.practical}
\begin{bchapter}
\bauthor{\bsnm{Greene}, \binits{D.}},
\bauthor{\bsnm{Cunningham}, \binits{P.}}:
\bctitle{Practical solutions to the problem of diagonal dominance in kernel
  document clustering}.
In: \bbtitle{Proceedings of the 23rd International Conference on Machine
  Learning}.
\bsertitle{ICML '06},
pp. \bfpage{377}--\blpage{384}.
\bpublisher{ACM},
\blocation{New York, NY, USA}
(\byear{2006}).
doi:\doiurl{10.1145/1143844.1143892}
\end{bchapter}
\endbibitem

\bibitem{phoneme}
\begin{botherref}
The phoneme database:
  \href{https://www.openml.org/d/1489}{https://www.openml.org/d/1489}, accessed
  1 feb 2019.
\end{botherref}
\endbibitem

\bibitem{Rice.2015.expected}
\begin{botherref}
\oauthor{\bsnm{Rice}, \binits{S.H.}}:
The expected value of the ratio of correlated random variables.
Texas Tech University
(2015)
\end{botherref}
\endbibitem

\end{thebibliography}

\newcommand{\BMCxmlcomment}[1]{}

\BMCxmlcomment{

<refgrp>

<bibl id="B1">
  <title><p>Classification of Content and Users in BitTorrent by
  Semi-supervised Learning Methods</p></title>
  <aug>
    <au><snm>Avrachenkov</snm><fnm>K.</fnm></au>
    <au><snm>Gon{\c c}alves</snm><fnm>P.</fnm></au>
    <au><snm>Legout</snm><fnm>A.</fnm></au>
    <au><snm>Sokol</snm><fnm>M.</fnm></au>
  </aug>
  <source>International Wireless Communications and Mobile Computing Conference
  (3rd International Workshop on Traffic Analysis and Classification)</source>
  <publisher>Cyprus</publisher>
  <pubdate>2012</pubdate>
  <note>Best paper award</note>
</bibl>

<bibl id="B2">
  <title><p>Soft-supervised Learning for Text Classification</p></title>
  <aug>
    <au><snm>Subramanya</snm><fnm>A</fnm></au>
    <au><snm>Bilmes</snm><fnm>J</fnm></au>
  </aug>
  <source>Proceedings of the Conference on Empirical Methods in Natural
  Language Processing</source>
  <publisher>Stroudsburg, PA, USA: Association for Computational
  Linguistics</publisher>
  <series><title><p>EMNLP '08</p></title></series>
  <pubdate>2008</pubdate>
  <fpage>1090</fpage>
  <lpage>-1099</lpage>
</bibl>

<bibl id="B3">
  <title><p>Compact Graph based Semi-Supervised Learning for Medical Diagnosis
  in Alzheimer’s Disease</p></title>
  <aug>
    <au><snm>{Zhao}</snm><fnm>M.</fnm></au>
    <au><snm>{Chan}</snm><fnm>R. H. M.</fnm></au>
    <au><snm>{Chow}</snm><fnm>T. W. S.</fnm></au>
    <au><snm>{Tang}</snm><fnm>P.</fnm></au>
  </aug>
  <source>IEEE Signal Processing Letters</source>
  <pubdate>2014</pubdate>
  <volume>21</volume>
  <issue>10</issue>
  <fpage>1192</fpage>
  <lpage>1196</lpage>
</bibl>

<bibl id="B4">
  <title><p>{BGP Zombies: an Analysis of Beacons Stuck Routes}</p></title>
  <aug>
    <au><snm>Fontugne</snm><fnm>R</fnm></au>
    <au><snm>Bautista</snm><fnm>E</fnm></au>
    <au><snm>Petrie</snm><fnm>C</fnm></au>
    <au><snm>Nomura</snm><fnm>Y</fnm></au>
    <au><snm>Abry</snm><fnm>P</fnm></au>
    <au><snm>Gon{\c c}alves</snm><fnm>P</fnm></au>
    <au><snm>Fukuda</snm><fnm>K</fnm></au>
    <au><snm>Aben</snm><fnm>E</fnm></au>
  </aug>
  <source>{PAM 2019 - 20th Passive and Active Measurements Conference}</source>
  <publisher>Puerto Varas, Chile</publisher>
  <pubdate>2019</pubdate>
  <fpage>1</fpage>
  <lpage>13</lpage>
  <url>https://hal.inria.fr/hal-01970596</url>
</bibl>

<bibl id="B5">
  <title><p>PageRank as a discrete Green's function</p></title>
  <aug>
    <au><snm>Chung</snm><fnm>F.</fnm></au>
  </aug>
  <source>Geometry and Analysis I ALM</source>
  <pubdate>2010</pubdate>
</bibl>

<bibl id="B6">
  <title><p>{Mean Field Analysis of Personalized PageRank with Implications for
  Local Graph Clustering}</p></title>
  <aug>
    <au><snm>Avrachenkov</snm><fnm>K</fnm></au>
    <au><snm>Kadavankandy</snm><fnm>A</fnm></au>
    <au><snm>Litvak</snm><fnm>N</fnm></au>
  </aug>
  <source>{Journal of Statistical Physics}</source>
  <publisher>{Springer Verlag}</publisher>
  <pubdate>2018</pubdate>
  <volume>173</volume>
  <issue>3-4</issue>
  <fpage>895</fpage>
  <lpage>916</lpage>
  <url>https://hal.inria.fr/hal-01936016</url>
</bibl>

<bibl id="B7">
  <title><p>Characterization of Tail Dependence for In-Degree and
  PageRank</p></title>
  <aug>
    <au><snm>Litvak</snm><fnm>N</fnm></au>
    <au><snm>Scheinhardt</snm><fnm>W</fnm></au>
    <au><snm>Volkovich</snm><fnm>Y</fnm></au>
    <au><snm>Zwart</snm><fnm>B</fnm></au>
  </aug>
  <source>Algorithms and Models for the Web-Graph</source>
  <publisher>Berlin, Heidelberg: Springer Berlin Heidelberg</publisher>
  <editor>Avrachenkov, Konstantin and Donato, Debora and Litvak, Nelly</editor>
  <pubdate>2009</pubdate>
  <fpage>90</fpage>
  <lpage>-103</lpage>
</bibl>

<bibl id="B8">
  <title><p>Four Cheeger-type Inequalities for Graph Partitioning
  Algorithms</p></title>
  <aug>
    <au><snm>Chung</snm><fnm>F.</fnm></au>
  </aug>
  <source>Proceedings of ICCM</source>
  <pubdate>2007</pubdate>
</bibl>

<bibl id="B9">
  <title><p>Pagerank Based Clustering of Hypertext Document
  Collections</p></title>
  <aug>
    <au><snm>Avrachenkov</snm><fnm>K</fnm></au>
    <au><snm>Dobrynin</snm><fnm>V</fnm></au>
    <au><snm>Nemirovsky</snm><fnm>D</fnm></au>
    <au><snm>Pham</snm><fnm>SK</fnm></au>
    <au><snm>Smirnova</snm><fnm>E</fnm></au>
  </aug>
  <source>Proceedings of the 31st Annual International ACM SIGIR Conference on
  Research and Development in Information Retrieval</source>
  <pubdate>2008</pubdate>
  <fpage>873</fpage>
  <lpage>-874</lpage>
</bibl>

<bibl id="B10">
  <title><p>Distributing Antidote Using PageRank Vectors</p></title>
  <aug>
    <au><snm>Graham</snm><fnm>FC</fnm></au>
    <au><snm>Horn</snm><fnm>P</fnm></au>
    <au><snm>Tsiatas</snm><fnm>A</fnm></au>
  </aug>
  <source>Internet Mathematics</source>
  <pubdate>2009</pubdate>
  <volume>6</volume>
  <fpage>237</fpage>
  <lpage>254</lpage>
</bibl>

<bibl id="B11">
  <title><p>Using PageRank to Locally Partition a Graph</p></title>
  <aug>
    <au><snm>Andersen</snm><fnm>R</fnm></au>
    <au><snm>R. K. Chung</snm><fnm>F</fnm></au>
    <au><snm>J. Lang</snm><fnm>K</fnm></au>
  </aug>
  <source>Internet Mathematics</source>
  <pubdate>2007</pubdate>
  <volume>4</volume>
  <fpage>35</fpage>
  <lpage>64</lpage>
</bibl>

<bibl id="B12">
  <title><p>Detecting Sharp Drops in PageRank and a Simplified Local
  Partitioning Algorithm</p></title>
  <aug>
    <au><snm>Andersen</snm><fnm>R</fnm></au>
    <au><snm>Chung</snm><fnm>F</fnm></au>
  </aug>
  <source>Theory and Applications of Models of Computation</source>
  <publisher>Berlin, Heidelberg: Springer Berlin Heidelberg</publisher>
  <editor>Cai, Jin-Yi and Cooper, S. Barry and Zhu, Hong</editor>
  <pubdate>2007</pubdate>
  <fpage>1</fpage>
  <lpage>-12</lpage>
</bibl>

<bibl id="B13">
  <title><p>Learning with Local and Global Consistency</p></title>
  <aug>
    <au><snm>Zhou</snm><fnm>D</fnm></au>
    <au><snm>Bousquet</snm><fnm>O</fnm></au>
    <au><snm>Lal</snm><fnm>TN</fnm></au>
    <au><snm>Weston</snm><fnm>J</fnm></au>
    <au><snm>Sch\"{o}lkopf</snm><fnm>B</fnm></au>
  </aug>
  <source>Advances in Neural Information Processing Systems 16</source>
  <editor>S. Thrun and L. K. Saul and B. Sch\"{o}lkopf</editor>
  <pubdate>2004</pubdate>
  <fpage>321</fpage>
  <lpage>-328</lpage>
  <url>http://papers.nips.cc/paper/2506-learning-with-local-and-global-consistency.pdf</url>
</bibl>

<bibl id="B14">
  <title><p>Spectral Clustering and Transductive Learning with Multiple
  Views</p></title>
  <aug>
    <au><snm>Zhou</snm><fnm>D</fnm></au>
    <au><snm>Burges</snm><fnm>CJC</fnm></au>
  </aug>
  <source>Proceedings of the 24th International Conference on Machine
  Learning</source>
  <publisher>New York, NY, USA: ACM</publisher>
  <series><title><p>ICML '07</p></title></series>
  <pubdate>2007</pubdate>
  <fpage>1159</fpage>
  <lpage>-1166</lpage>
</bibl>

<bibl id="B15">
  <title><p>Generalized Optimization Framework for Graph-based Semi-supervised
  Learning</p></title>
  <aug>
    <au><snm>Avrachenkov</snm><fnm>K.</fnm></au>
    <au><snm>Gon{\c c}alves</snm><fnm>P.</fnm></au>
    <au><snm>Mishenin</snm><fnm>A.</fnm></au>
    <au><snm>Sokol</snm><fnm>M.</fnm></au>
  </aug>
  <source>SIAM Data Mining</source>
  <pubdate>2012</pubdate>
</bibl>

<bibl id="B16">
  <title><p>Semi-supervised Learning by Higher Order Regularization</p></title>
  <aug>
    <au><snm>Zhou</snm><fnm>X</fnm></au>
    <au><snm>Belkin</snm><fnm>M</fnm></au>
  </aug>
  <source>Proceedings of the Fourteenth International Conference on Artificial
  Intelligence and Statistics</source>
  <publisher>Fort Lauderdale, FL, USA: PMLR</publisher>
  <editor>Geoffrey Gordon and David Dunson and Miroslav Dudík</editor>
  <series><title><p>Proceedings of Machine Learning
  Research</p></title></series>
  <pubdate>2011</pubdate>
  <volume>15</volume>
  <fpage>892</fpage>
  <lpage>-900</lpage>
  <url>http://proceedings.mlr.press/v15/zhou11b/zhou11b.pdf</url>
</bibl>

<bibl id="B17">
  <title><p>Diffusion and Clustering on Large Graphs</p></title>
  <aug>
    <au><snm>Tsiatas</snm><fnm>A</fnm></au>
  </aug>
  <source>PhD thesis</source>
  <publisher>University of California at San Diego</publisher>
  <pubdate>2012</pubdate>
</bibl>

<bibl id="B18">
  <title><p>Graph-based Semi-supervised Learning Methods and Quick Detection of
  Central Nodes</p></title>
  <aug>
    <au><snm>Sokol</snm><fnm>M.</fnm></au>
  </aug>
  <source>PhD thesis</source>
  <publisher>Universit{\'e} de Nice, Ecole Doctorale STIC</publisher>
  <pubdate>2014</pubdate>
</bibl>

<bibl id="B19">
  <title><p>Fractional dynamics on networks: Emergence of anomalous diffusion
  and L\'evy flights</p></title>
  <aug>
    <au><snm>P\'erez Riascos</snm><fnm>A</fnm></au>
    <au><snm>Mateos</snm><fnm>J</fnm></au>
  </aug>
  <source>Physical Review E</source>
  <pubdate>2014</pubdate>
  <volume>90</volume>
  <fpage>032809</fpage>
</bibl>

<bibl id="B20">
  <title><p>Fractional graph-based semi-supervised learning</p></title>
  <aug>
    <au><snm>{de Nigris}</snm><fnm>S.</fnm></au>
    <au><snm>{Bautista}</snm><fnm>E.</fnm></au>
    <au><snm>{Abry}</snm><fnm>P.</fnm></au>
    <au><snm>{Avrachenkov}</snm><fnm>K.</fnm></au>
    <au><snm>{Goncalves}</snm><fnm>P.</fnm></au>
  </aug>
  <source>2017 25th European Signal Processing Conference (EUSIPCO)</source>
  <pubdate>2017</pubdate>
  <fpage>356</fpage>
  <lpage>360</lpage>
</bibl>

<bibl id="B21">
  <title><p>{L{\'e}vy Flights for Graph Based Semi-Supervised Classification.
  }</p></title>
  <aug>
    <au><snm>Bautista</snm><fnm>E</fnm></au>
    <au><snm>De Nigris</snm><fnm>S</fnm></au>
    <au><snm>Abry</snm><fnm>P</fnm></au>
    <au><snm>Avrachenkov</snm><fnm>K</fnm></au>
    <au><snm>Gon{\c c}alves</snm><fnm>P</fnm></au>
  </aug>
  <source>{26th colloquium GRETSI}</source>
  <publisher>Juan-Les-Pins, France</publisher>
  <series><title><p>GRETSI, 2017 - Proceeding of the 26th
  colloquium</p></title></series>
  <pubdate>2017</pubdate>
</bibl>

<bibl id="B22">
  <title><p>Distributed Signal Processing via Chebyshev Polynomial
  Approximation</p></title>
  <aug>
    <au><snm>{Shuman}</snm><fnm>D. I.</fnm></au>
    <au><snm>{Vandergheynst}</snm><fnm>P.</fnm></au>
    <au><snm>{Kressner}</snm><fnm>D.</fnm></au>
    <au><snm>{Frossard}</snm><fnm>P.</fnm></au>
  </aug>
  <source>IEEE Transactions on Signal and Information Processing over
  Networks</source>
  <pubdate>2018</pubdate>
  <volume>4</volume>
  <issue>4</issue>
  <fpage>736</fpage>
  <lpage>751</lpage>
</bibl>

<bibl id="B23">
  <title><p>Reconstruction and estimation in the planted partition
  model</p></title>
  <aug>
    <au><snm>Mossel</snm><fnm>E</fnm></au>
    <au><snm>Neeman</snm><fnm>J</fnm></au>
    <au><snm>Sly</snm><fnm>A</fnm></au>
  </aug>
  <source>Probability Theory and Related Fields</source>
  <pubdate>2015</pubdate>
  <volume>162</volume>
  <issue>3</issue>
  <fpage>431</fpage>
  <lpage>-461</lpage>
</bibl>

<bibl id="B24">
  <title><p>Phase transitions in semisupervised clustering of sparse
  networks</p></title>
  <aug>
    <au><snm>Zhang</snm><fnm>P</fnm></au>
    <au><snm>Moore</snm><fnm>C</fnm></au>
    <au><snm>Zdeborova</snm><fnm>L</fnm></au>
  </aug>
  <source>Physical review. E, Statistical, nonlinear, and soft matter
  physics</source>
  <pubdate>2014</pubdate>
  <volume>90</volume>
</bibl>

<bibl id="B25">
  <title><p>Comparison of the predicted and observed secondary structure of T4
  phage lysozyme</p></title>
  <aug>
    <au><snm>Matthews</snm><fnm>B.W.</fnm></au>
  </aug>
  <source>Biochimica et Biophysica Acta (BBA) - Protein Structure</source>
  <pubdate>1975</pubdate>
  <volume>405</volume>
  <issue>2</issue>
  <fpage>442</fpage>
  <lpage>451</lpage>
  <url>http://www.sciencedirect.com/science/article/pii/0005279575901099</url>
</bibl>

<bibl id="B26">
  <title><p>Gradient-based learning applied to document recognition</p></title>
  <aug>
    <au><snm>{Lecun}</snm><fnm>Y.</fnm></au>
    <au><snm>{Bottou}</snm><fnm>L.</fnm></au>
    <au><snm>{Bengio}</snm><fnm>Y.</fnm></au>
    <au><snm>{Haffner}</snm><fnm>P.</fnm></au>
  </aug>
  <source>Proceedings of the IEEE</source>
  <pubdate>1998</pubdate>
  <volume>86</volume>
  <issue>11</issue>
  <fpage>2278</fpage>
  <lpage>2324</lpage>
</bibl>

<bibl id="B27">
  <title><p>Distinctive Descriptions for Face Processing.</p></title>
  <aug>
    <au><snm>Hond</snm><fnm>D</fnm></au>
    <au><snm>Spacek</snm><fnm>L</fnm></au>
  </aug>
  <source>BMVC</source>
  <editor>Clark, Adrian F.</editor>
  <pubdate>1997</pubdate>
</bibl>

<bibl id="B28">
  <title><p>Practical Solutions to the Problem of Diagonal Dominance in Kernel
  Document Clustering</p></title>
  <aug>
    <au><snm>Greene</snm><fnm>D</fnm></au>
    <au><snm>Cunningham</snm><fnm>P</fnm></au>
  </aug>
  <source>Proceedings of the 23rd International Conference on Machine
  Learning</source>
  <publisher>New York, NY, USA: ACM</publisher>
  <series><title><p>ICML '06</p></title></series>
  <pubdate>2006</pubdate>
  <fpage>377</fpage>
  <lpage>-384</lpage>
</bibl>

<bibl id="B29">
  <title><p>The Phoneme Database:
  \href{https://www.openml.org/d/1489}{https://www.openml.org/d/1489}, Accessed
  1 Feb 2019.</p></title>
</bibl>

<bibl id="B30">
  <title><p>The expected value of the ratio of correlated random
  variables</p></title>
  <aug>
    <au><snm>Rice</snm><fnm>S. H.</fnm></au>
  </aug>
  <source>Texas Tech University</source>
  <pubdate>2015</pubdate>
</bibl>

</refgrp>
} 







\end{backmatter}
\end{document}